%% file: ElfringLauga_arxiv2.tex
\documentclass[aps,pre,reprint,onecolumn,notitlepage,groupedaddress]{revtex4-1}
\usepackage{epsfig,graphics,amssymb,amsmath,subeqnarray,color}
\usepackage[sfdefault=cmbr]{isomath}

\input{definitions}

\makeindex

\begin{document}

\title{Theory of locomotion through complex fluids}
\author{Gwynn J Elfring\footnote{Email: gelfring@mech.ubc.ca}}
\affiliation{
Department of Mechanical Engineering, 
University of British Columbia}
\author{Eric Lauga\footnote{Email: e.lauga@damtp.cam.ac.uk}}
\affiliation{
Department of Applied Mathematics and Theoretical Physics, 
University of Cambridge}
\date{\today}
\begin{abstract}
Microorganisms such as bacteria often swim in fluid environments that cannot be classified as Newtonian. Many biological fluids  contain polymers or other heterogeneities which may yield complex rheology. For a given set of boundary conditions on a moving organism, flows can be substantially different in complex fluids, while non-Newtonian stresses can alter the gait of the microorganisms themselves. Heterogeneities in the fluid may also be characterized by length scales on the order of the organism itself leading to additional dynamic  complexity. In this chapter we present a  theoretical overview  of small-scale locomotion in complex fluids with a focus on recent efforts quantifying the impact of non-Newtonian rheology on swimming microorganisms.\\

Chapter appears in \textit{Complex Fluids in Biological Systems}, Saverio E. Spagnolie (Ed.), Springer (2015).
\end{abstract}

\maketitle

\section{Introduction}\label{sec:1}

Many microorganisms swim through fluids that display non-Newtonian characteristics. For example, as spermatozoa make their journey through the female reproductive tract they encounter several complex fluids including glycoprotein-based cervical mucus in the cervix \cite{katz78}, mucosal epithelium inside the fallopian tubes, and an actin-based viscoelastic gel outside the ovum \cite{dunn76, suarez06}. These complex fluids often have dramatic effects on the locomotion of microorganisms. Human sperm flagella beat with higher frequency but smaller amplitude and wavelength in cervical mucus compared to semen. This results in roughly the same swimming speed but along straighter paths due to a reduction of wobbling and end effects \cite{katz78}. Sperm hyperactivation (larger amplitude, asymmetric beating patterns) increases the ability to penetrate viscoelastic fluids \cite{suarez91,suarez92}. Experimental evidence suggests therefore that spermatozoa both passively and actively modulate their swimming kinematics due to the presence of non-Newtonian stresses. In contrast, the bacterium \textit{Helicobacter pylori} actively modulates the viscoelastic properties of its environment in order to move \cite{celli09}. \textit{H. pylori} lives in the human stomach and produces urease which leads to a drastic reduction of viscoelastic moduli, allowing the bacterium to swim freely. 

In order to understand these, and related, effects, one must develop a theory for locomotion in complex fluids. The equations of motion governing the flow of most non-Newtonian fluids are nonlinear and hence classical Stokes flow methods involving the superposition of fundamental solutions are invalid. As a result, useful properties which constrain locomotion in a Newtonian fluid, such as the kinematic reversibility of the field equations,   break down in viscoelastic fluids. The presence of time-dependent stresses, normal stress differences, and shear-dependent material functions in complex fluids are able to fundamentally alter the physics of locomotion \cite{purcell77, lauga09b}. In this chapter  we present a very general overview of the theoretical framework  used to describe the effect of complex fluids on the locomotion of microorganisms. In section \ref{sec:2} we elucidate the mathematical framework used to study locomotion in fluids and review well-established principles governing swimming in Newtonian fluids. In section \ref{sec:3} complex constitutive relations are introduced and considered in this framework. Section \ref{sec:4} presents analytical results obtained for geometrically simple model swimmers in complex fluids, and comparisons with numerical simulation and theory are made. Finally we close this chapter by offering our perspective on the direction of research in this area in section \ref{sec:5}.

\section{Locomotion in fluids}\label{sec:2}\index{Locomotion}
Experience may furnish the reader with intuition on swimming in fluids but as illustrated in the previous chapter of this book (Ch.~7), locomotion in fluids is quite different for humans than it is for microorganisms. In this section we present a mathematical definition of locomotion in fluids, elucidate what it means to swim in a fluid if one is very small, and demonstrate consequences if the fluid is Newtonian.

\subsection{Boundary motion}

\begin{figure}[t]
\centering
\includegraphics[width=0.4\textwidth]{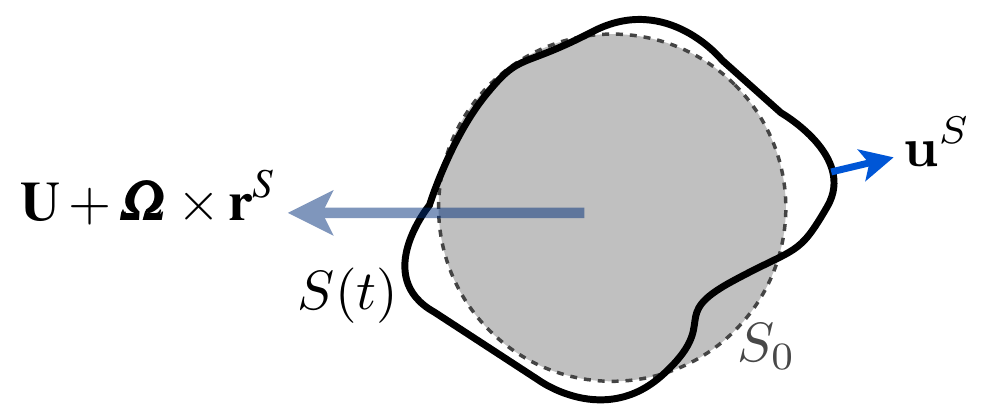}
\caption{Schematic representation of a general swimmer, adapted from 
Ref.~\cite{lauga09}. A swimmer is defined as  body with whose surface deforms in time thereby effecting an instantaneous rigid-body translation, $\bU$, and rotation, $\bOmega$.}
\label{swimmer}
\end{figure}

In order to swim, a body undergoes (periodic) changes in its surface $S(t)$ (see 
Fig.~\ref{swimmer}). When in a fluid, this surface deformation leads to stresses exerted from the fluid on the body and, in general, motion. Periodic deformations may be described as deviations from a reference surface $S_0$. The position, $\bx^S$, of a point on the surface of a swimmer $S(t)$, may be decomposed as
\begin{align}
\bx^S(t) = \bx_0(t)+\br^S(t),
\end{align}
where $\bx_0$ is a body-fixed position (the center of mass). The swimming gait of the body is defined using a body-fixed frame as follows
\begin{align}
\br^S(t) = \bTheta(t)\cdot\br(t),
\end{align}
where the rotation operator, $\bTheta$, orients the reference frame as 
\begin{equation}
\frac{\d \Theta_{ik}}{\d t}\Theta_{jk} =-\epsilon_{ijk}\Omega_k,\end{equation}
in which  $\bOmega$ is the angular velocity. Upon differentiation of the position of a point on the body we obtain the velocity
\begin{align}\label{swimmerbc}
\frac{\d \bx^S}{\d t}=\frac{\d \bx_0}{\d t}+\frac{\d\bTheta}{\d t}\cdot\bTheta^\top\cdot\bTheta\cdot\br+\bTheta\cdot\frac{\d\br}{\d t}=\bU+\bOmega\times\br^S+\bu^S.
\end{align}

Stresses imparted on the body by the fluid may lead to an instantaneous rigid-body translation, $\bU$, and/or rotation, $\bOmega$, due to Newton's second law. Mathematically, one can freely move between the lab frame and the body-fixed frame, the sole difference being that the rigid-body motion of the body is either reflected on the body or at infinity. In non-inertial frames, additional inertial forces have to also be considered in general. 

If a typical time scale of the periodic deformation is $\omega^{-1}$, for a body of size $L$, and density $\rho_b$, in a fluid of dynamic viscosity $\eta$, and density $\rho$, the Stokes number $\rho_b \omega L^2/\eta$, determines the magnitude of inertial forces versus viscous forces on the body. Similarly the ratio of gravitational body forces to viscous fluid forces is given an Archimedes number, $\Ar=gL(\rho_b -\rho)/\eta\omega$. For microorganisms of sufficiently small length scales and which are close to density matched with the surrounding fluid, we can ignore the inertial and body force terms and thus the body is instantaneously  force- and torque-free, 
\begin{align}
\bF = \int_S \bn\cdot\bsigma \d S &=\bzero,\\
\bL = \int_S \br^S\times(\bn\cdot\bsigma) \d S &=\bzero,
\end{align}
where the surface $S$ is implicitly a function of time and the normal to the surface $\bn$ points into the fluid. The stress tensor $\bsigma$ can be decomposed into an isotropic and deviatoric part $\bsigma = -p\bI + \btau$ (see Ch.~2 in this book for more details). We assume here the fluid is incompressible, $\bnabla\cdot\bu=0$, and that the Reynolds number, $\Re = \rho\omega L^2/\eta$, is small and so any fluid parcel is in instantaneous mechanical equilibrium
\begin{align}\label{cauchy}
\bnabla\cdot\bsigma=\bzero.
\end{align}

In a Newtonian fluid the deviatoric stress is linearly proportional to the strain-rate tensor
\begin{align}
\btau = \eta\bgammad = \eta(\bnabla\bu+\bnabla\bu^\top),
\end{align}
where $^\top$ is the transpose. Upon substition of this constitutive equation into Eq.~\ref{cauchy} we have, together with incompressibility, the Stokes equations
\begin{align}
\eta\nabla^2\bu &= \bnabla p\\
\bnabla\cdot\bu &=0.
\end{align}
The linearity of the Stokes equations allows the superposition of solutions and so we can think of the problem of swimming as the sum of two, conceptually simpler problems. In the first, a body is held fixed and so the boundary motion is due only to the swimming gait $\bu(\bx^S)=\bu^S$. Because $\bU=\bzero$ and $\bOmega=\bzero$, there may arise a non-zero hydrodynamic force, $\bF_1$, and torque, $\bL_1$, on the body (indicated with a 1 subscript). In the second problem, conversely, an undeforming swimmer with the same instantaneous shape is subjected to rigid-body motion and so $\bu(\bx^S) = \bU+\bOmega\times\br^S$. This rigid-body motion leads to drag forces (indicated by a 2 subscript). The force and the torque on the body may be written as a linear equation
\begin{align}
\begin{pmatrix}
\bF_2 \\
\bL_2
\end{pmatrix}
=-
\begin{pmatrix}
\bR_{FU} & \bR_{F\Omega}\\
\bR_{LU} & \bR_{L\Omega}
\end{pmatrix}\cdot
\begin{pmatrix}
\bU \\
\bOmega
\end{pmatrix},
\end{align}
where the resistance tensors, $\bR_{FU}$, $\bR_{F\Omega}$, $\bR_{LU}$ and $\bR_{L\Omega}$ connect the kinematics to the force and torque. We write this relationship more compactly in terms of six-dimensional tensors as $\tF_2 = -\tR \cdot\tU$. Naturally, since swimming is force and torque free, we must have $\tF_1+\tF_2 = \tzero$, (essentially thrust balances drag) and so the rigid-body-motion arising during free swimming is simply
\begin{align}\label{prob1}
\tU = \tR^{-1}\cdot\tF_1.
\end{align}
This simple form belies the fact that we still must solve for both the hydrodynamic forces, $\tF_1$, and the rigid-body resistance tensor, $\tR$, at each instant for a deforming body, a considerable task even in a Newtonian fluid.

The rate of work (power) expended by a swimming organism by deforming its surface in time in the fluid is instantaneously equal to the total energy dissipation rate in the fluid exterior to $S$,
\begin{align}
P&=\int_S -\bn\cdot\bsigma\cdot\bu\d S=\int_V \btau:\bnabla\bu \d V.
\end{align}
The contraction $\bA:\bB \equiv A_{ij}B_{ji}$ in this chapter. If the fluid is Newtonian we may write
\begin{equation}
P = \frac{\eta}{2}\int_{V}\bgammad:\bgammad\d V.
\end{equation}

\subsection{The reciprocal theorem}\index{Reciprocal theorem}
Solving for the motion of the swimmer is complicated by the fact that the boundary conditions are not fully prescribed but must satisfy the integral constraints of force- and torque-free motion for all times. Stone and Samuel showed in Ref.~\cite{stone96} that determining the rigid-body motion of the free swimmer, $\bU$ and $\bOmega$, may be greatly simplified by appealing to the reciprocal theorem \cite{happel65}.

We denote $\bu$ and $\bsigma$ as the velocity field and its associated stress tensor for a force- and torque-free swimmer of surface $S$, while $\buh$ and $\bsigmah$ the velocity field and its associated stress tensor for a body of the same instantaneous shape subject to rigid-body translation and rotation with speeds $\bUh$ and $\bOmegah$.  Each fluid is in mechanical equilibrium $\bnabla\cdot\bsigma=\bnabla\cdot\bsigmah=\bzero$ so by the equality of virtual powers
\begin{align}
\int_V \bnabla\cdot\bsigmah\cdot\bu \d V=\int_V \bnabla\cdot\bsigma\cdot \buh \d V=0,
\end{align}
where the volume of fluid $V$ is external to the surface $S$ with normal $\bn$ into the fluid \footnote{There was a sign error due to the normal in earlier versions of this work}. Invoking the divergence theorem we obtain
\begin{align}\label{11}
\int_S \bn\cdot\bsigmah\cdot\bu\d S+\int_V \bsigmah:\bnabla\bu\d V=\int_S \bn\cdot\bsigma\cdot\buh\d S +\int_V \bsigma:\bnabla\buh\d V.
\end{align}
Because the swimmer is force- and torque-free, the first term on the right-hand side of Eq.~\eqref{11} is zero and hence so is the second term on the right-hand side by construction,
\begin{align}\label{first}
\int_V \bsigma:\bnabla\buh\d V &= 0.
\end{align}
The first term on the left-hand side of Eq.~\eqref{11} may be expanded by using the boundary motion on $S$ from Eq.~\eqref{swimmerbc}, meanwhile the stress tensor $\bsigma = -p\bI +\btau$, and so assuming the fluids to be incompressible,
\begin{align}
\bFh\cdot\bU+\bLh\cdot\bOmega&=-\int_S \bn\cdot\bsigmah\cdot\bu^S\d S-\int_V \btauh:\bnabla\bu\d V.
\end{align}
Here $\bFh$ and $\bLh$ represent the force and torque resulting from the rigid-body motion of $S$. As we shall show, if both fluids are Newtonian, then the last term vanishes and one is left simply with the swimming kinematics $\bU$ and $\bOmega$ as a function of known quantities $\bu^S$ and the auxiliary rigid-body problem. For the purposes of this work we will always assume that the fluid in the rigid-body problem is Newtonian, a dramatic simplification that is without penalty. Taking $\btauh=\etah\bgammadh$ we can write
\begin{align}
\bFh\cdot\bU+\bLh\cdot\bOmega=-\int_S \bn\cdot\bsigmah\cdot\bu^S\d S -\etah\int_V\bgammad:\bnabla\buh\d V,
\end{align}
where we have used the identity $\bgammadh:\bnabla\bu = \bgammad:\bnabla\buh$. Due to the linearity of the Stokes equations we may write $\buh = \tGh\cdot\tUh$, $\bsigmah = \tTh\cdot\tUh$ while $\tFh = -\tRh\cdot\tUh$ (using six-dimensional tensors for compactness, and so $\tTh$ is $[3\times3]\times6$). The resistance tensor takes the form
\begin{align}
\tRh =-\int_S
\begin{bmatrix}
\bn\cdot\tTh\\
\br\times\lb(\bn\cdot\tTh\rb)
\end{bmatrix}
\d S.
\end{align}
Substituting into the reciprocal theorem yields
\begin{align}
-\tU\cdot\tRh\cdot\tUh=-\int_S \bu^S\bn:\tTh\cdot\tUh\d S -\etah\int_V\bgammad:\bnabla\tGh\cdot\tUh\d V.
\end{align}
where $\bu^S\bn:\tTh\cdot\tUh = u^S_jn_i\textsf{T}_{ijk}\textsf{U}_k$. Discarding the arbitrary vector $\tUh$, and using the symmetry of the resistance tensor $\tRh$ \cite{brenner63}, we finally arrive at a general integral theorem for swimming
\begin{align}\label{motiongeneral}
\tU=\tRh^{-1}\cdot\lb[\int_S \bu^S\bn:\tTh\d S +\etah\int_V \bgammad:\bnabla\tGh\d V\rb].
\end{align}
The volume integral in  Eq.~\eqref{motiongeneral} only contributes if the fluid in the swimming problem is {\it not} Newtonian and hence is a measure of the modification of the swimming dynamics due to the presence of non-Newtonian stresses.

There remain two difficulties however: the first is that the surface $S(t)$ is changing and so the auxiliary rigid-body problem must be solved for all possible shapes; the second is the unknown integrand in the volume integral. We show below that both difficulties may be tackled by taking a perturbative approach. 

Note that the reciprocal theorem above may be extended  to $N$ swimmers by taking $S=\bigcup_\alpha S^\alpha$ which gives 
\begin{align}
\tU=\tRh^{-1}\cdot\lb[\lb(\sum_\alpha\int_{S^\alpha} \bu^S\bn:\tTh\d S\rb) +\etah\int_V \bgammad:\bnabla\tGh\d V\rb],
\end{align}
except now the tensors are $6N$ in size where $N$ is the number of bodies.

\subsection{Swimming in Newtonian fluids}
When swimming in a Newtonian fluid, Eq.~\eqref{first} dictates that
\begin{align}
\int_V \bgammad:\bnabla\buh\d V &= 0
\end{align}
and so the velocity of the swimmer, as given by Eq.~\eqref{motiongeneral}, reduces simply to
\begin{align}\label{locomotion}
\tU
= \tRh^{-1}\cdot\lb[\int_S\bu^S\bn:\tTh\d S\rb].
\end{align}
This is a marked simplification because now the swimming motion, $\tU = [\bU \ \bOmega]^\top$, can be resolved without the knowledge of the entire flow field $\bu$, rather one must simply ascertain the solution of the auxiliary field $\buh$ and its associated stress tensor. Swimming in a Newtonian fluid is therefore reduced to solving, pointwise, the stress on a body of the same shape undergoing instantaneous rigid-body translation. We observe that the term in the brackets is simply the hydrodynamic force and torque exerted by the fluid on the swimmer if it was held instantaneously in place (see Eq.~\eqref{prob1} for comparison).
In practice one must solve the auxiliary problem of rigid-body translation and rotation in a Newtonian fluid for each shape of $S(t)$ for all time $t$ at a resolution desired for integrating the instantaneous velocity. Note that because of the relationship between $\tRh$ and $\tTh$, Eq.~\eqref{locomotion} implies that the locomotion of the body is independent of the viscosity of the fluid. As an example, if the body is a sphere of radius $a$, then the rigid-body problem is well known, the resistance matrix is diagonal and hence easily invertible, $\bRh_{FU}=6\pi\eta a\bI$, $\bRh_{L\Omega}=8\pi\eta a^3\bI$, $\bRh_{F\Omega}=\bzero$ and $\bRh_{LU}=\bzero$, while $\bn\cdot\tTh = -\frac{3\eta}{2a}[\bI \ \ 2\bXi]$ where $\Xi_{ij}=\epsilon_{ijk}r_k$. With these the swimming speed for a sphere with only tangential deformations is given by
\begin{align}
\tU = -\frac{1}{4\pi a^2}\int_S
\begin{bmatrix}
\bI\\
\frac{3}{2a^2}\bXi^\top
\end{bmatrix}\cdot\bu^S\d S,
\end{align}
as shown by Stone and Samuel \cite{stone96}.

Swimming gaits are typically periodic in time (with period $T$), and often we are only interested in the steady swimming speed given by a time-average of Eq.~\eqref{locomotion}
\begin{align}\label{netlocomotion}
\lb<\tU\rb>
= \lb<\tRh^{-1}\cdot\lb[\int_S \bu^S\bn:\tTh\d S\rb]\rb>,
\end{align}
where $\langle \tU\rangle \equiv T^{-1}\int_T \tU \d t$.

\subsubsection{The scallop theorem}\label{scalloptheorem}\index{Scallop theorem}
Because the Stokes equations are linear and independent of time, time only enters the problem as a parameter in the boundary conditions and the locomotion is instantaneously linear in $\bu^S$ (Eq.~\ref{locomotion}). This has two profound implications on locomotion in Newtonian fluids. The first is that the rate of actuation of the boundaries is irrelevant to the distance traveled over a period -- in other words it does not matter if the actuation rate is fast or slow, over a period \cite{purcell77}. This is because re-parameterizing time $t'=f(t)$ changes the velocity $\bu^S \rightarrow \fd\bu'^S$ but also the interval $dt\rightarrow \fd^{-1}dt'$ to no net effect. This also implies that if a swimmer reverses its sequence of shapes after a period it goes back to where it started, regardless of the rate of motion. Finally if a swimmer goes through the same sequence of shapes whether forwards and backwards in time (time-reversible) then it will have no net motion. More precisely stated, if the deformation of the swimmer shape over a period, $t_2-t_0$, is such that there exists a  $t_1 \in (t_0,t_2)$ where the sequence of shapes after $t_1$ is exactly reversed, then the net motion is zero \cite{lauga09b}. We can show  this by invoking Eq.~\eqref{netlocomotion} as follows
\begin{align}
\lb<\tU\rb>&=\frac{1}{t_2-t_0}\int_{t_0}^{t_2}\int_{S(t)} \bu^S\bn:\tTh\cdot\tRh^{-1} \d S\d t, \\
&=\frac{1}{t_2-t_0}\lb[\int_{t_0}^{t_1} \int_{S(t)} \bu^S\bn:\tTh\cdot\tRh^{-1} \d S\d t +\int_{t_1}^{t_2}\int_{S(t')} \bu'^S\bn:\tTh\cdot\tRh^{-1} \d S\d t'\rb],\nonumber\\
&=\frac{1}{t_2-t_0}\lb[\int_{t_0}^{t_1} \int_{S(t)} \bu^S\bn:\tTh\cdot\tRh^{-1} \d S\d t+\int_{t_1}^{t_0}\int_{S(t)} \bu^S\bn:\tTh\cdot\tRh^{-1} \fd^{-1} \d S\fd\d t\rb],\nonumber\\
&=\tzero\nonumber.
\end{align}
This statement is often called the \textit{scallop theorem} because the physical actuation of a scallop was used by Purcell to first elucidate this principle \cite{purcell77} (for a detailed mathematical treatment see \cite{ishimoto12}). The scallop is an example of a swimmer with only one degree of freedom, namely its hinge. In general, if a body has only one degree of freedom it can only execute time-reversible motion if it is to perform a cyclical gait and thus no such body can swim in a Newtonian fluid in the absence of inertia \cite{lauga09}. 

This impediment means often that the simplest designs of swimmers in the inertial realm, for example objects with a rigid flapping tail, cannot locomote at small scales in Newtonian fluids. We will see below that in non-Newtonian fluids, the mathematical details which lead to the scallop theorem, namely  linearity and an independence of time, are no longer present and time-reversible swimmers can generally self-propel.

\subsection{Small-amplitude motion}\label{smallamplitude}
When a swimming motion results from small deformations of a body from a reference surface $S_0$, as is the case for ciliated organisms \cite{blake71}, one can describe this motion theoretically by writing   
\begin{equation}
\br^S = \bTheta\cdot\lb[\br_0 + \Delta\br(\br_0,t)\rb]
\end{equation}
where $\Delta\br$ is small. The position of a point on the reference surface $\bx^{S_0} =\bx_0 +\bTheta\cdot\br_0$ and $\bx^S-\bx^{S_0} = \bTheta\cdot\Delta\br$. Such a gait presents several mathematical advantages for calculating the swimming kinematics. In general, in order to use the reciprocal formulation one must know the solution to the auxiliary problem $\bsigmah$ for all $S(t)$, which is typically impractical for a nontrivial gait. When $S(t)$ deviates only slightly from $S_0$ we can, through Taylor series expansions, recast the problem onto $S_0$ \cite{felderhof94a,felderhof94b}. As the shape of $S_0$ is invariant in time we then need only the resolution of a single auxiliary problem. Furthermore, as  discussed below, by posing the problem as a perturbation expansion in $\epsilon$ we are able to tackle  the constitutive relationships in a systematic fashion.

\subsubsection{Recasting the problem onto $S_0$}
The swimming gait, $\bu^S(\bx^S)$, represents motion of the material points, $\bx^S$, on the swimmer surface. To represent the boundary condition on a surface, $S_0$ which is not material, we Taylor expand the flow field $\bu$,
\begin{align}
\bu(\bx^S)&=\bu(\bx^{S_0})+(\bx^S-\bx^{S_0})\cdot\lb.\bnabla\bu\rb|_{\bx_0^S}+\Of\lb(\lvert\bx^S-\bx^{S_0}\rvert^2\rb),\\
\Rightarrow\bu(\bx^{S_0})&=\bU+\bOmega\times\br^S+\bu^S-(\bx^S-\bx^{S_0})\cdot\lb.\bnabla\bu\rb|_{\bx^{S_0}}+\Of\lb(\lvert\bx^S-\bx^{S_0}\rvert^2\rb),\label{velonbody}
\end{align}
where we have used $\bu(\bx^S)=\bU+\bOmega\times\br^S+\bu^S$ and where $|_{\bx^{S_0}}$ means that derivatives are evaluated at $\bx^{S_0}$. This may be rewritten as
\begin{equation}\label{bcsmall}
\bu(\bx^{S_0},t) = \bU +\bOmega\times\br^{S_0}+\bu^{S_0}.
\end{equation}
The first two terms in Eq.~\eqref{velonbody} represent rigid-body rotation of the undeforming surface $S_0$, and the remaining term is the boundary condition that one must impose on $S_0$ to obtain the appropriate solution on $S$. We refer to this term, $\bu^{S_0}$, as the swimming gait defined on $S_0$ where $\br^{S_0} = \bTheta\cdot\br_0$ and
\begin{equation}
\bu^{S_0}=\bu^S+\bOmega\times(\br^S-\br^{S_0})-(\bx^S-\bx^{S_0})\cdot\lb.\bnabla\bu\rb|_{\bx^{S_0}}+\Of\lb(\lvert\bx^S-\bx^{S_0}\rvert^2\rb).
\end{equation}
This formulation leads to a swimming problem defined entirely on $S_0$ which satisfies the correct boundary conditions on $S$.

In order to solve for the swimming kinematics, we then apply the reciprocal theorm on $S_0$ (namely assuming a rigid-body translation and rotation of $S_0$ for the auxiliary field). The motion of a point on $S_0$ in the swimming problem is given by $\bu(\bx^{S_0}) =\bU + \bOmega\times\br^{S_0}+\bu^{S_0}$ while for rigid-body motion $\buh(\bx^{S_0}) = \bUh + \bOmegah\times\br^{S_0}$. In the swimming problem the force and torque on $S$ are both zero and because the fluid stress in the volume between $S_0$ and $S$ is divergence free, $\bnabla\cdot\bsigma=\bzero$, the total force and torque on $S_0$ must also be zero. Applying the integral theorem for swimming, Eq.~\eqref{motiongeneral}, on the surface $S_0$ we get
\begin{align}\label{mgrecast}
\tU=\tRh^{-1}\cdot\lb[\int_{S_0} \bu^{S_0}\bn:\tTh\d S +\etah\int_{V_0} \bgammad:\bnabla\tGh\d V\rb].
\end{align}
The great benefit of this approach is that one need only resolve the rigid-body translation for a single shape, that of the undeforming surface $S_0$. Note that we do not get something for nothing. In particular the swimming gait on $S_0$, $\bu^{S_0}$, depends on gradients of the (unknown) flow field $\bu$ and the rotation rate of the swimmer $\bOmega$. However, because these terms are $\Of\lb(\lvert\bx^S-\bx^{S_0}\rvert\rb)$, upon a perturbation expansion they will vanish to leading order.

\section{Locomotion in non-Newtonian fluids}\label{sec:3}
The motion of a swimming body in a viscous fluid is characterized, in general, by the integral relationship given by Eq.~\eqref{motiongeneral}. In a non-Newtonian fluid, the expression for the deviatoric stress, $\btau$, is not linear in the strain-rate tensor, $\bgammad$, and hence the term in brackets on the right-hand side of Eq.~\eqref{motiongeneral} will not vanish even as we continue to take the auxiliary fluid to be Newtonian $\btauh = \eta \bgammadh$. Let us assume for the moment, for illustrative purposes, that the stress tensor may be decomposed into a Newtonian contribution and an additional non-Newtonian part, $\btau = \eta \bgammad +\bA(\bx,t)$. With this form of constitutive equation, Eq.~\eqref{first} yields
\begin{align}
\int_V \bgammad:\bnabla\buh\d V &= -\frac{1}{\eta}\int_V \bA:\bnabla\buh\d V
\end{align}
and so substitution into Eq.~\eqref{motiongeneral} leads to
\begin{align}\label{motiongeneral2}
\tU=\tRh^{-1}\cdot\lb[\int_S \bu^S\bn:\tTh\d S -\frac{\etah}{\eta}\int_V \bA:\bnabla\tGh\d V\rb].
\end{align}
or, substitution into \eqref{mgrecast}, for the problem recast onto $S_0$, yields
\begin{align}\label{mgrecast2}
\tU=\tRh^{-1}\cdot\lb[\int_{S_0} \bu^{S_0}\bn:\tTh\d S -\frac{\etah}{\eta}\int_{V_0} \bA:\bnabla\tGh\d V\rb].
\end{align}

Swimming in a non-Newtonian fluid involves the solution of a second integral which is, in general, a function of the complex flow field and may depend on the history of the deformations $S(t)$. For example, if the sequence of shapes is time-reversible, then there may still exist net locomotion due to that non-Newtonian integral, a breakdown of the scallop theorem \cite{lauga11}.

The dissipation due to swimming would likewise be modified by nonlinearities as
\begin{align}
P = \frac{\eta}{2}\int_{V}\bgammad:\bgammad\d V+\int_{V}\bA:\bnabla\bu\d V.
\end{align}

In order to derive a more precise statement about swimming in non-Newtonian fluids we have to specify a constitutive relationship that gives rise to non-Newtonian behavior. For modeling purposes we assume here that the deviatoric stress tensor can be decomposed into a sum of  relaxation modes $j$, $\btau = \sum \btau^{(j)}$. We write the relationship between each stress mode and the velocity field very generally as
\begin{align}
\Af_j\btau^{(j)}=\eta_j\Bf_j\bgammad+\bN_j(\bu,\btau^{(j)}),\label{constrelgen}
\end{align}
where $\eta_j$ is the zero-shear-rate viscosity for the $j$-th mode, $\Af_j$ and $\Bf_j$ are linear operators in time and $\bN_j$ is a symmetric tensor which depends nonlinearly on the velocity and stress and represents the transport and stretching of the polymeric microstructure by the flow. This general constitutive relationship includes, in particular, all classical  models of polymeric fluids \cite{lauga09}.

From this point forward it can be very difficult to make analytical progress in large part because of the presence of the nonlinear operators $\bN_j$. One way to make progress in light of this difficulty is to consider  the constitutive relationship perturbatively. In this manner at each order, the nonlinear terms are functions of the previous order solutions only. We detail this approach below.

\subsection{Small-amplitude perturbations}
As shown in section \ref{smallamplitude}, when the deformation of the body is small, $\epsilon \ll 1$, one can recast the problem onto a body whose shape is not deforming, $S_0$ (recall that  $\epsilon$ is a dimensionless measure of the amplitude deformation of the surface $S(t)$). By employing perturbation expansions, the nonlinear non-Newtonian constitutive equations are linearized order-by-order facilitating analytical solution.

Expanding all fields formally in a regular perturbation series, e.g. $\bu = \sum_m \epsilon^m\bu_m$, the boundary condition, Eq.~\eqref{bcsmall}, becomes at each order $m$
\begin{align}
\bu_m(\bx_0^S) = \bU_m +\bOmega_m\times\br^{S_0}+\bu^{S_0}_m,
\end{align}
where $\bu^{S_0}_1 =\bu^S_1$. The constitutive relation at each order $m$ becomes
\begin{align}
\Af_j\btau_m^{(j)}&=\eta_j\Bf_j\bgammad_m+\bN_m^{(j)}[\bu_1,...,\bu_{m-1}],
\end{align}
where $\bN$ is a functional of previous order solutions, for example $\bN_1^{(j)}=\bzero$ and $\bN_2^{(j)}\equiv\bN_2^{(j)}[\bu_1]$.

\subsubsection{Fourier series}
Given a swimmer with a time-periodic swimming gait, it is a reasonable assumption to write the flow and stress fields as  time periodic and expand them in a Fourier series in time as
\begin{align}
\bu = \sum_n\bu^{(n)}e^{in\omega t}.
\end{align}
As we shall describe below, this assumption means we neglect  the influence of a particular initial stress state in the fluid, but is suitable for determining the steady swimming speed of a microorganism, and all harmonic oscillations around it. The constitutive relationship for each Fourier mode, $n$, is
\begin{align}
\btau^{(j,n)}&=\eta_j\frac{B_j(n)}{A_j(n)}\bgammad^{(n)}+\frac{1}{1+A_j(n)}\bN^{(j,n)},\nonumber\\
&=\eta^*_j(n)\bgammad^{(n)}+\bA^{(j,n)},
\end{align}
where $A_j(n)$ and $B_j(n)$ are the characteristic polynomials of the differential operators (i.e $e^{-in\omega t}\Af_j[e^{in\omega t}]$) while $A_j(0)=B_j(0)=1$. Summing over all relaxation modes, $j$, we then have
\begin{align}
\btau^{(n)}=\eta^*(n)\bgammad^{(n)}+\bA^{(n)}.\label{constrel}
\end{align}
For each Fourier mode we thus have a linear response with complex viscosity, $\eta^*(n)$, and a nonlinear term which depends on the solutions at previous orders.  

We may now decompose the boundary conditions into Fourier modes and solve for the flow field order-by-order in $\epsilon$ using the aforementioned small-amplitude expansion about the static surface $S_0$. Upon substitution of \eqref{constrel} into the reciprocal relationship for swimming, Eq.~\eqref{mgrecast2}, we obtain, order-by-order,
\begin{align}\label{modeswimming}
\tU_m^{(n)}=\tRh^{-1}\cdot\lb[\int_{S_0} \bu_m^{S_0,(n)}\bn:\tTh\d S -\frac{\etah}{\eta^*(n)}\int_{V_0} \bA_m^{(n)}:\bnabla\tGh\d V\rb],
\end{align}
while the mean swimming speed is given by the zeroth Fourier mode, $\lb<\bU\rb>=\bU^{(0)}$, and thus satisfies
\begin{align}\label{modemean}
\lb<\tU_m\rb>=\tRh^{-1}\cdot\lb[\int_{S_0} \lb<\bu_m^{S_0}\rb>\bn:\tTh\d S -\frac{\etah}{\eta^*(n)}\int_{V_0} \lb<\bA_m\rb>:\bnabla\tGh\d V\rb],
\end{align}
where we have used the notation that $\eta_0 = \eta^*(0)$ is the zero-shear-rate viscosity. The values of both $\etah$ and $\eta_0$ do not affect $\tU$ and so we may set $\etah=\eta_0$ for the sake of convenience.

At leading order, $m=1$, $\bA_1=\bzero$ and so the swimming speed, as shown by Eq.~$\eqref{modeswimming}$, is the Newtonian one  for all times. This is expected because we simply have a set of Stokes equations for each Fourier mode
\begin{align}
\bnabla p_1^{(n)}=\eta^*(n)\bnabla^2\bu_1^{(n)}.
\end{align}
The velocity field is independent of the viscosity and identical to the Newtonian solution. Furthermore, because the kinematics of the swimmer are periodic with zero mean, by Eq.~\eqref{modemean} the swimmer will have zero mean translation or rotation to leading order. Using the integral equation to determine the kinematics of the body, $\bU_1^{(n)}$ and $\bOmega_1^{(n)}$, we may then proceed to solve the Stokes equations and obtain the entire flow field, $\bu_1$.

At second order, $m=2$, one need not solve for the full velocity field, $\bu_2$, in order to find the mean swimming velocity $\lb<\bU_2\rb>$. One must simply compute the mean of the nonlinear tensor $\bA_2^{(0)}[\bu_1]$ and the mean of the gait at second order,
\begin{align}
\lb<u^{S_0}_2\rb>=\lb<\bu^S_2+\bOmega_1\times\bx_1^S(\bx_0^S,t)-\bx_1^S(\bx_0^S,t)\cdot\bnabla\lb.\bu_1\rb|_{\bx^{S_0}}\rb>,
\end{align}
both of which depend only on the first order solution, $\bu_1$.

Now consider a swimmer with a time-reversible gait. Such a swimmer will  have zero net motion in a Newtonian fluid as demonstrated in section \ref{scalloptheorem}, and hence its velocity in a complex fluid will be entirely determined by non-Newtonian stresses. Consider for example a sphere with tangential surface motion only, a model known as a squirmer. The shape is not deforming, and if the gait is time-reversible then $\lb<\bu^{S_0}\rb>=\bzero$. The leading-order swimming speed is given by the integral
\begin{align}
\lb<\tU\rb> = -\epsilon^2\frac{\etah}{\eta_0}\int_{V_0}\lb<\bA_2\rb>:\bnabla\tGh\cdot\tRh^{-1}\d V+\Of(\epsilon^4),
\end{align}
where for a sphere of radius $a$
\begin{align}
\tGh\cdot\tRh^{-1} =\frac{1}{8\pi\eta}
\begin{bmatrix}
\lb(1+\frac{a^2}{6}\nabla^2\rb)\bG \ & \ \frac{1}{\lb|\bx\rb|^3}\bXi
\end{bmatrix}.
\end{align}
and $\bG = \frac{1}{\lb|\bx\rb|}\lb(\bI +\frac{\bx\bx}{\lb|\bx\rb|^2}\rb)$ is the Oseen tensor.
We see that there is an $\Of(\epsilon^2)$, strictly non-Newtonian swimming speed which arises from the nonlinear terms in a given constitutive relationship, $\bA_2$. Because we chose a gait which does not locomote in a Newtonian fluid, any net motion is then a measure of the non-Newtonian rheology of the fluid. For a generic gait  which achieves locomotion in a Newtonian fluid, there is instead a non-Newtonian correction to the swimming speed at quadratic order in amplitude.
\color{black}

\subsubsection{Linear viscoelasticity}
Here we address an important point on locomotion in linearly viscoelastic fluids. As we have seen, the swimming speed of a microorganism is Newtonian at linear order in amplitude while non-Newtonian effects do not appear until quadratic order. If the fluid is linearly viscoelastic then $\bA_m=\bzero$ at all orders and hence the fluid yields no change in swimming speed from that of a Newtonian fluid for prescribed kinematics, as found in Ref.~\cite{fulford98}.

It is also typical for organisms executing time-periodic gaits to exhibit $\epsilon\rightarrow-\epsilon$ symmetry. In such a case the Newtonian swimming speed itself is at least quadratic in amplitude and hence it is desirable to keep nonlinearities in the constitutive relation which would emerge at the same order as the swimming speed itself \cite{smith09}.

\subsubsection{Transients}
Specifying periodicity in time, as done above,  neglects any transient regime in which  stresses and velocities  evolve from an initial condition,  $\{\btau(0),\bu(0)\}$, to a periodic steady state. This simplification is desirable when we are concerned with determining the time-averaged steady-state swimming speed of an organism. Swimming organisms are however intrinsically unsteady organisms in the sense that while the swimming gait may be periodic over a short period, the swimmers may stop and start or change direction, as exemplified by bacteria executing \textit{run-and-tumble} motion \cite{berg72}. Here we examine briefly the effects on locomotion of an arbitrary initial stress state. We note that we will still assume that the swimmer is instantaneously force free, in other words that the inertial time scale is still much smaller than the relevant relaxation time scale. 

Taking the Laplace transform of the constitutive relation, 
Eq.~\eqref{constrelgen}, we obtain
\begin{align}
\widetilde{\Af_j\btau_m^{(j)}}=\eta_j\widetilde{\Bf_j\bgammad_m}+\widetilde{\bN_m^{(j)}},
\end{align}
where the tilde indicates a unilateral Laplace transform $\tilde{f}(s) = \int_{0}^{\infty}f(t) e^{-st}\d t$. Rearranging and summing over all relaxation modes, $j$, we may write
\begin{align}\label{stresslaplace}
\btaut_m=\eta^*(s)\bgammadt_m+\bAt_m+\bBt_m,
\end{align}
where
\begin{align}
\eta^*(s) &= \sum_j \eta_j\frac{B_j(s)}{A_j(s)},\\
\bAt_m &= \sum_j\frac{1}{A_j(s)}\widetilde{\bN_m^{(j)}}.
\end{align}
The tensor $\bBt$ represents the effect of the initial condition on the stress. For example if our fluid of interest a single-mode Boger fluid \cite{james09}, then for the Oldroyd-B equations have $\Af = 1+\lambda_1 \partial_t$ while $\Bf = 1+\lambda_2 \partial_t$, where $\lambda_1$ is the relaxation time and $\lambda_2$ is the retardation time, and hence
\begin{align}
\eta^*(s) &= \eta_0\frac{1+\lambda_2 s}{1+\lambda_1 s},\\
\bAt_m &= \frac{1}{1+\lambda_1 s}\widetilde{\bN_m},\\
\bBt_m &= \frac{\lambda_1}{1+\lambda_1 s}\lb[\btau_m(t=0)-\eta_0(\lambda_2/\lambda_1)\bgammad_m(t=0)\rb].
\end{align}

In order to understand the effect the constitutive equation \eqref{stresslaplace} has on locomotion we again appeal to the reciprocal theorem for swimming
\begin{align}
\tU_m=\tRh^{-1}\cdot\lb[\int_{S_0} \bu_m^{S_0}\bn:\tTh\d S -\etah\int_{V_0} \Lf^{-1}\lb(\frac{\bAt_m+\bBt_m}{\eta^*(s)}\rb):\bnabla\tGh\d V\rb],
\end{align} 
where $\Lf^{-1}$ indicates an inverse Laplace transform. Again we see that we need not know the solution for the flow field at order $m$ to determine the swimming kinematics at that order. The nonlinear term, $\bA_m$, is a function of the flow field at previous orders while the tensor $\bB_m$ consists entirely of the initial conditions. 

At leading order the nonlinear contribution vanishes, $\bA_1=\bzero$. In an Oldroyd-B fluid the contribution at leading order may thus be written as
\begin{align}
\tU_1=\tRh^{-1}\cdot\lb[\int_{S_0} \bu_1^{S_0}\bn:\tTh\d S -\frac{\etah}{\eta_0}e^{-t/\lambda_2}\int_{V_0} \lb(\frac{\lambda_1}{\lambda_2}\btau_1(0)-\eta_0\bgammad_1(0)\rb):\bnabla\tGh\d V\rb].
\end{align} 
We see that the influence of the initial condition decays exponentially on the retardation time scale of the fluid. That time scale needs then to be compared with the other relevant time scales in each specific swimming problem.

\subsection{Slowly varying flows}
We have demonstrated the utility of small-amplitude deformations as a method to probe the nonlinear effects of a particular non-Newtonian fluid in the context of small-scale locomotion. A particular benefit of such an approach is that the actuation time scales of the microorganisms can be arbitrary in comparison to the relaxation time scales of the non-Newtonian medium in which they are moving \cite{lauga09}. 

Conversely, one may be interested in gaits which do not display small-amplitude motions but instead lead to slowly varying flows. In such cases, one can resort to the use of the second-order fluid model which describes the first non-Newtonian behavior in an expansion of stress in strain rate. In the weakly nonlinear regime the deviatoric stress for almost all complex fluids can be represented by 
\begin{align}
\btau=\eta \bgammad - \frac{1}{2}\Psi_1\stackrel{\triangledown}{\bgammad}+\Psi_2\bgammad\cdot\bgammad,
\end{align}
where $\Psi_1$ and $\Psi_2$ are the first and second normal stress coefficients \cite{bird87a,khair10}, and so
\begin{align}
\bA = - \frac{1}{2}\Psi_1\stackrel{\triangledown}{\bgammad}+\Psi_2\bgammad\cdot\bgammad.
\end{align}
If we scale strain rates $\bgammad=\omega\bgammad'$ and stresses $\btau=\eta\omega\btau'$, where $\omega$ is the characteristic actuation frequency of the body, then we have in dimensionless form (primes indicate dimensionless quantities)
\begin{align}
\btau' = \bgammad'-\De\lb(\upc{\bgammad'}+B\bgammad'\cdot\bgammad'\rb).
\end{align}
The Deborah number, $\De = \omega \Psi_1/2\eta$, is the ratio of the relaxation time scale of the fluid compared to the time scale of actuation and $B=-2\Psi_2/\Psi_1\ge0$. 

If  we now assume a regular perturbation expansion of the flow field in Deborah number, $\bu' =\bu'_0 +\De\bu'_1+...$, then at leading order we have simply a Newtonian fluid
\begin{align}
\btau_0' = \bgammad_0',
\end{align}
while at first order we have
\begin{align}
\btau_1' = \bgammad_1' - \lb(\upc{\bgammad_0'}+B\bgammad_0'\cdot\bgammad_0'\rb)\equiv\bgammad_1'+\bA'_1[\bu_0'].
\end{align}

\subsubsection{Locomotion}
Using the above constitutive relation in the integral theorem for swimming \eqref{motiongeneral} we obtain
\begin{align}
\tU'=\tRh'^{-1}\cdot\lb[\int_{S'} \bu'^S\bn:\tTh'\d S' -\De\int_{V'} \bA'_1:\bnabla'\tGh'\d V'\rb]+\Of(\De^2).
\end{align}
Hence in order to compute the $\Of(\De)$ correction to the swimming speed we need only solve for the Newtonian flow field, $\bu_0(\bx)$. 

Consider the counterrotation of two connected but different axisymmetric bodies (two unequal spheres for example). Such a swimmer will not locomote in Stokes flow due to the kinematic reversibility of the field equations but may have net motion in a second-order fluid due to the $\Of(\De)$ term. Because we explicitly chose a gait which would not swim in a Newtonian fluid, the rigid-body motion of the swimmer is given by the non-Newtonian contribution only,
\begin{align}
\tU'=-\De\int_{V'} \bA'_1:\bnabla'\tGh'\cdot\tRh'^{-1}\d V'+\Of(\De^2).
\end{align}
The swimming speed is at best linear in Deborah number. The dimensional swimming speed scales thus as
$\bU \sim \omega L \De$, where $L$ is the typical length scale. Note this is invariant under a reversal of actuation $\omega\rightarrow-\omega$ (because $\De\sim \omega$), as expected due to the axisymmetry of the body. The Newtonian problem may not be available analytically but one could obtain the solution numerically using the boundary integral method, for instance.

Now due to the symmetry of this body, we expect the only rigid-body motion to be translation along the axis connecting the spheres. For this reason it is unnecessary to compute the full resistance tensor. In order to find the swimming speed one needs only to solve the auxiliary problem of rigid-body translation of the same body in the direction of swimming and the integration reduces to
\begin{align}\label{slowspeed}
U'=-\De R'^{-1}_{FU}\int_{V'} \bA'_1:\bnabla'\buh'\d V'+\Of(\De^2),
\end{align}
where $\buh'=\buh/\hat{U}$ is the dimensionless flow field due to rigid-body translation along the axis of rotation.

\subsubsection{Rheology}
In the previous section we were able to construct locomotion which depended on the presence of non-Newtonian stresses. Rather than calculating the swimming speed for a known constitutive equation, we may use this technique instead to infer unknown rheological properties, namely the normal stress coefficients $\Psi_1$ and $\Psi_2$, of a fluid through experimental measurement of the swimming kinematics of deforming bodies \cite{pak12} based on theory proposed by Khair and Squires \cite{khair10}. This idea is summarized below.

One can recast Eq.~\eqref{slowspeed} as a linear equation in the normal stress coefficients
\begin{align}
U &= \frac{\omega^2 L}{\eta}\lb[C_1\Psi_1+C_2\Psi_2\rb],
\end{align}
where
\begin{align}
C_1 &= \frac{1}{2}\Rh_{FU}^{-1}\int_{V'}\upc{\bgammad'_0}:\bnabla'\buh'\d V',\\
C_2 &= -\Rh_{FU}^{-1}\int_{V'}\bgammad'_0\cdot\bgammad'_0:\bnabla'\buh'\d V',
\end{align}
are so-called coupling coefficients, functions of the zeroth order solution, independent of the non-Newtonian properties of the flow and hence functions of geometry alone.

Since typically $\Psi_1 \gg \lb|\Psi_2\rb|$, a reasonable approximation of the first normal stress coefficient would be
\begin{align}
\Psi_1 \approx \frac{\eta U}{C_1\omega^2L}\cdot
\end{align}
By measuring swimming kinematics in a non-Newtonian fluid (one which does not display net motion in a Newtonian fluid) we may thus obtain a good approximation of the first normal stress coefficient. 

In order to unambiguously determine both the first and second normal stress coefficients, two independent measurements must be made. One may, for example, devise another independent swimmer or one could measure kinematics other than a swimming speed. An example of the latter would be to measure the relative displacement of two equal counterrotating spheres. Experimentally this could be accomplished by an axial coupling that transmits torque but allows free translation. In such a case it is helpful to decompose the body as $S=S_1\cup S_2$ where $S_1$ and $S_2$ are the two sphere surfaces. By symmetry we expect only a relative displacement of the spheres $\bU^R=\bU^1-\bU^2=2\bU^1$, equal and opposite, along the axis of rotation (with resistance to this motion, defined by $\Rh_{FU}^R$) and so this is the only rigid-body motion that needs resolution. By kinematic reversibility we know the rotation of the spheres produces no net displacement in a Newtonian fluid so the reciprocal theorem yields directly
\begin{align}
U'^{R}= -\De\frac{1}{\Rh'^{R}_{FU}}\int_{V'} \bA_1':\bnabla'\buh'\d V'+\Of(\De^2).
\end{align}
where $\buh'= \buh/(\hat{U}^R/2)$ is the dimensionless flow field due to rigid-body translation of two equal spheres away from one another with relative speed $\Uh^R$. With this second kinematic measurement one has two equations for the unknown normal stress coefficients of the fluid. We can write this simply in matrix form,
\begin{align}
\begin{pmatrix}
U\\
U^R
\end{pmatrix}
=\frac{L\omega^2}{\eta}
\begin{pmatrix}
C_1 & C_2\\
C_1^R & C_2^R
\end{pmatrix}
\begin{pmatrix}
\Psi_1\\
\Psi_2
\end{pmatrix}.
\end{align}
The normal stress coefficients are obtained by inverting the matrix of coupling coefficients, $\bC$. Practically, one may manipulate the geometries in the experiment so as to minimize the condition number of $\bC$ and hence the potential error in the values of the normal stress coefficients obtained through the measurements of $U$ and $U^R$ which are subject to experimental error  \cite{khair10}.

\begin{figure}
\centering
\includegraphics[width=0.65\textwidth]{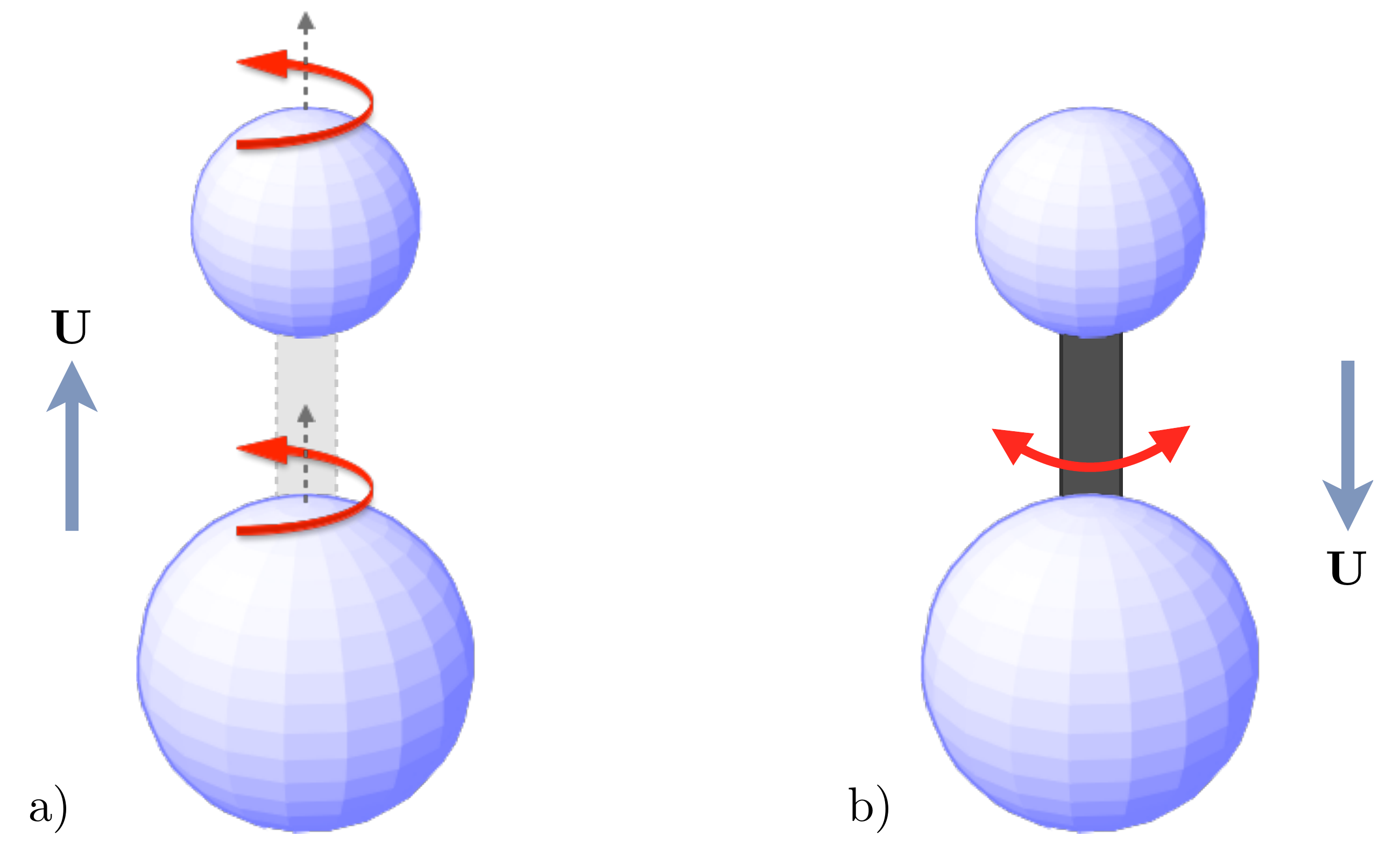}
\caption{(a) Two unequal spheres rotating together as a rigid body will move in the direction of the smaller sphere due to an imbalance of hydrodynamic interactions driven by viscoelastic stresses \cite{pak12}. (b) Experiments on two rigidly connected spheres rotating about an out-of-plane axis leads to net viscoelastic locomotion in the direction of the larger sphere \cite{keim12}.}
\label{motors}
\end{figure}

Although we have emphasized here experimental observation of the kinematics of force- and torque-free bodies as means to decipher the normal stress coefficients,  one could alternatively directly measure the forces on bodies which move with prescribed kinematics. In the method first proposed by Khair and Squires \cite{khair10}, one would pull two equal spheres in the direction joining their two centers, $\bU_\parallel$, in a first experiment and in a direction orthogonal to the line joining their centers, $\bU_\perp$, in a second experiment, each time measuring the force on each sphere. The difference between the forces on each sphere is then a measure of the non-Newtonian stresses, because in each case these forces would be identical in a Newtonian fluid. The suggested experimental method would be to trap two particles in a dual optical trap, then translate the bulk fluid uniformly and measure the difference in the trapping force on each probe \cite{khair10}.

Alternatively Pak and Lauga proposed a setup involving force-free, but not torque-free spheres. In their model, two spheres are externally  rotated together as a rigid body about the axis connecting their centers (see Fig.~\ref{motors}a). In that case the spheres will impart a net torque onto the fluid  and are thus not  strictly swimmers. Nevertheless, because the spheres impart no net force on the fluid, the equations for translation precisely mimic those presented in this section for a force- and torque-free swimmer.  Physically, the direction in which such an object moves can be understood by means of the hoop stresses generated along curved streamlines. A secondary, purely elastic flow, is created by each rotating sphere, contracting in along the equator of each sphere and flowing out of the poles. Because the spheres are unequal in size,  hydrodynamic interactions due to this secondary flow are unbalanced leading to propulsion in the direction of the smallest sphere. The same mechanism causes a net drift when a single rotating sphere is placed near a wall in a viscoelastic fluid. The rotating sphere drives fluid radially out in the direction of its poles and if a no-slip wall is placed opposite of one pole, the sphere is driven away from the wall. 

A similar object, two unequal spheres connected by a rigid rod, was used in experiments by Keim \textit{et al.} \cite{keim12}. In that case the spheres oscillate together as a rigid body about an axis orthogonal to the line connecting their centers (see Fig.~\ref{motors}b). This motion is time-reversible and hence does not yield time-averaged locomotion in a Newtonian fluid, yet in a viscoelastic fluid there is net migration. These experiments also demonstrated that a wall acts as a symmetry-breaking mechanism for the propulsion of a dimer with equal spheres. Analytical studies have also indicated that time-reversible motions of anchored bodies (flapping a rigid rod for instance) may pump fluid \cite{normand08,pak10} in a viscoelastic fluid when such net flow is not possible in a Newtonian fluid due to the scallop theorem.

\section{Infinite models}\label{sec:4}
In the previous section we looked at a formulation for the swimming speed based on a modification of the reciprocal theorem for locomotion in a viscoelastic fluid. Historically the simplest possible models for understanding swimmers in Newtonian fluids have been infinite models with reduced dimensionality, in particular the canonical Taylor swimming sheet \cite{taylor51}. Here again, the reciprocal theorem may be used to solve for the swimming speed of a two-dimensional sheet and yield insight into the effects of complex fluids interacting with nontrivial boundary actuation.

The analysis of the correction to the swimming speed of the swimming sheet due to a non-Newtonian fluid was first performed by Chaudhury using a second-order fluid (or Rivlin-Erickson fluid of grade 2) \cite{chaudhury79}, and later by Sturges for a second-order fluid of grade $N$ \cite{sturges81}. In both cases no change in swimming speed was observed  in the zero Reynolds number limit. The analysis was then performed for a variety of nonlinear non-Newtonian constitutive equations by Lauga \cite{lauga07}. There, it  was shown that, for fixed swimming kinematics, the swimming speed for the swimming sheet systematically decreases compared to the  Newtonian case for all Oldroyd-type fluids to leading order in small-amplitude waveforms. Subsequently the same result was obtained for planar undulating filaments \cite{fu07} and helical waves \cite{fu09} in Oldroyd-B fluids.

In the following we give an overview of the use of the reciprocal theorem to obtain, in a direct fashion, the influence of complex flows on the swimming sheet in a  general waveform. We also derive viscoelastic corrections to the swimming speed of a general sheet near walls.

\subsection{Taylor swimming sheet}
\begin{figure}[t]
\centering
\includegraphics[width=0.7\textwidth]{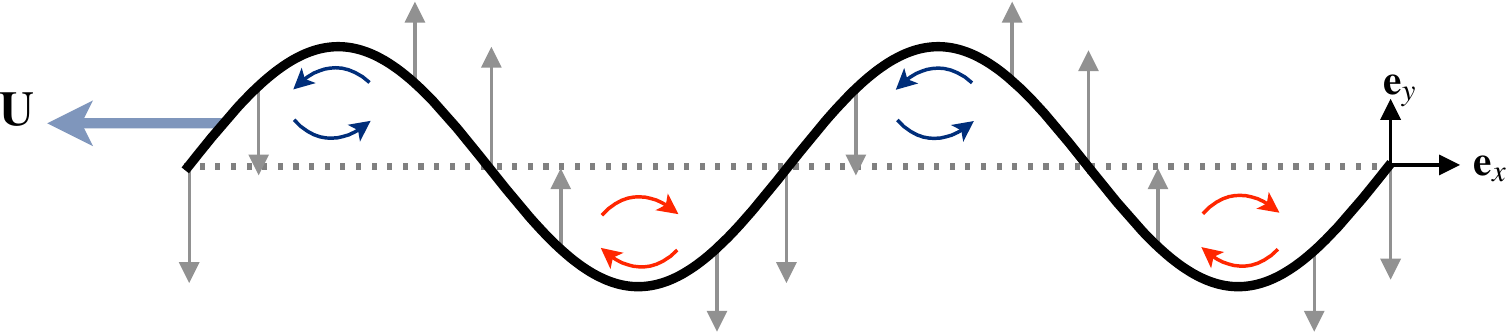}
\caption{A sinusoidal swimming sheet propagates waves of transverse oscillation in a Newtonian fluid (adapted from 
Ref.~\cite{lauga09b}). The grey arrows indicate the transverse oscilations of the wave which travels to the right at speed $\omega k^{-1}$. The vorticity generated drives the sheet to the left at speed $U$.}
\label{sheetdouble}
\end{figure}

\index{Taylor swimming sheet}
Consider a two-dimensional sheet which propagates traveling waves. In a frame moving with the traveling waves the sheet has the static shape $y_1=a g(\xi)$ where $g$ is written very generally as
\begin{align}
g(\xi)=\sum_n c_n e^{in\xi},
\end{align}
with $a$ being the dimensional wave amplitude and $\xi=kx-\omega t$ with wavenumber $k$ and frequency $\omega$. In Fig. \ref{sheetdouble} we illustrate a sinusoidal swimming sheet which has only one mode $c_1 = -i/2$. Ostensibly the sheet generates vorticity which is oppositely signed from peak to trough, and thus for a  waveform  traveling to the right the fluid forces act to drive the sheet to the left \cite{lauga09b}.

The boundary conditions for an inextensible sinusoidal sheet were first described by Taylor \cite{taylor51} and catalogued in detail in 
Ref.~\cite{elfring11} for general waveforms. Following Taylor, we take the approach that the amplitude of transverse oscillations of the sheet is small and expand all fields a regular perturbation series in $ak=\epsilon \ll 1$. The boundary conditions are redefined onto the surface $y=0$ ($S_0$) through Taylor expansion. Finally we introduce a no-slip wall at $y=h$,  and by taking the limit $h\rightarrow \infty$ we will obtain swimming in an unbounded fluid. We  take the auxiliary flow to be simple shear flow between $y=0$ and $y=h$, for which the  reciprocal theorem simplifies dramatically to give
\begin{align}\label{speed1}
U =-u_{x}^{S_0,(0)}+\frac{1}{\eta_0 }\int_{0}^hA_{xy}^{(0)} \d y,
\end{align}
where $u_{x}^{S_0,(0)}=\lb<\bu^{S_0}\rb>\cdot\be_x$ while $A_{xy}^{(0)}=\lb<\bA\rb>:\be_y\be_x$ (with basis vectors as shown in Fig.~\ref{sheetdouble}). This framework can also be used for a sheet unequally spaced between two walls at $y=h_1$ and $y=-h_2$, In that case, due to the lack of symmetry both the upper and lower surfaces of the sheet and the volume of fluid above and below each surface must be accounted for and the reciprocal theorem then yields
\begin{align}\label{speed2}
U =\frac{1}{h_1+h_2}\lb[-h_2u^{S_1,(0)}_{x}-h_1u^{S_2,(0)}_{x}+\frac{1}{\eta_0}\lb(h_2\int_0^{h_1}A_{xy}^{(0)}\d y+h_1\int_{-h_2}^0A_{xy}^{(0)}\d y\rb)\rb].
\end{align}

Coming back to the one-wall case,  we consider as an example an Oldroyd-B fluid \cite{bird87a}\index{Oldroyd-B}. The complex viscosity is given by
\begin{align}
\eta^* = \eta_0\frac{1+in \omega \lambda_2}{1+in\omega\lambda_1}\cdot
\end{align}
At leading order in $\epsilon$, $\bA_1 =\bzero$ and $\btau_1^{(n)} = \eta^*(n)\bgammad_1^{(n)}$, while at second order
\begin{align}
\bA_2^{(0)}&=\sum_q\lb[\frac{\eta_0}{iq\omega}\lb(1-\frac{\eta^*(q)}{\eta_0}\rb)\left(\bnabla \bu_1^\text{(-q)T}\cdot\bgammad_1^{(q)}+\bgammad_1^{(q)}\cdot\bnabla\bu_1^{(-q)}-\bu_1^{(-q)}\cdot\bnabla\bgammad_1^{(q)}\right)\rb].
\end{align}
Following the framework described in section \ref{sec:3}, we seek a perturbative solution to the swimming speed as $U(\epsilon) = \sum \epsilon^mU_m$ and similarly expand all fields in powers of $\epsilon$. Referring to Eq.~\eqref{speed1} one can find immediately that $U_1 = 0$.

In order to obtain the swimming speed at quadratic order, $U_2$, we must find the nonlinear contribution at second order, $\bA_2$, the boundary conditions, $\bu_2^{S_0}$, and the full leading-order flow field, $\bu_1$, to obtain the swimming speed at second order.  The mean of the boundary condition on $S_0$ is given by
\begin{align}
\bu_2^{S_0,(0)}&=\omega k^{-1}\sum_n n^2\Upsilon(n h) c_nc_n^\dagger\be_x,
\end{align}
where $\Upsilon(x) = (\sinh^2 (x)+x^2)/(\sinh^2 (x)-x^2)$. In a Newtonian fluid the swimming speed is given directly as 
\begin{equation}
\bU_2 = -\lb<\bu_2^{S_0}\rb>,
\end{equation}
and in particular when $h\rightarrow \infty$ we have $\Upsilon \rightarrow 1$. For Taylor's swimming sheet ($c_1=-i/2$) we recover the classical result,
\begin{equation}
\bU = -\frac{1}{2}\omega k^{-1}\epsilon^2\be_x+\Of(\epsilon^4).
\end{equation}

Resolving $\bA_2$ in the  Oldroyd-B case we have
\begin{align}
\frac{1}{\eta_0}\int_0^h A_{2,xy}^{(0)}\d y &= \omega k^{-1}\sum_n \lb[1-\frac{\eta^*(n)}{\eta_0}\rb]n^2c_nc_n^\dagger\Upsilon(nh).
\end{align}
Here we see that the nonlinear forcing term, for each mode, is proportional to the boundary condition on $S_0$ at second order. With this result the swimming speed is expressed as
\begin{align}
U_2 &=-\omega k^{-1}\sum_n n^2c_nc_n^\dagger\Upsilon(nh)\frac{\eta^*(n)}{\eta_0}\cdot
\end{align}
Each component of the Newtonian swimming speed is therefore simply scaled by the dimensionless viscous modulus. Now if the sheet is unevenly spaced between two walls, at distances $h_1$ and $h_2$, an application of Eq.~\eqref{speed2} yields
\begin{align}
U_2 =-\omega k^{-1}\sum_n n^2c_nc_n^\dagger \frac{h_2\Upsilon(n h_1)+h_1\Upsilon(n h_2)}{h_1+h_2}\frac{\eta^*(n)}{\eta_0}\cdot
\end{align}

In both cases we see that the factor which differentiates the result from the Newtonian swimming speed, namely that which arises strictly from non-Newtonian stresses, is simply the dimensionless loss modulus for each mode
\begin{align}
\Rf\lb[\frac{\eta^*(n)}{\eta_0}\rb]=\frac{1+n^2\beta\De^2}{1+n^2\De^2},
\end{align}
where $\De=\omega\lambda_1$ and $\beta=\lambda_2/\lambda_1$, $\beta < 1$ as it indicates the ratio of the suspending solvent viscosity to total viscosity in the fluid. When $\De\ll 1$, the fluid is probed on time scales much longer than it takes  to relax, and the Newtonian behavior is recovered $\Rf[\eta^*/\eta_0]\rightarrow 1$. When the Deborah number is nonzero the swimming speed is always less than the Newtonian one. In particular in the $\De \gg 1$ limit, we obtain $\Rf[\eta^*/\eta_0]\approx \beta$ and the swimming speed is simply a factor of the Newtonian swimming speed, $U=\beta U_N$.

The results obtained above indicate that in a Oldroyd-B fluid, each Fourier mode in the swimming speed is  rescaled by the  respective (dimensionless) viscosity for that mode. It may appear obvious that the linear response of the fluid should affect the swimming speed but it is not. If the fluid were simply linearly viscoelastic there would be no change in the swimming speed from that of a Newtonian fluid. It is thus the nonlinear response that leads to change in the locomotion speed. As the flow field scales  linearly with the sheet amplitude, the first nonlinear correction (the tensor $\bA_2$) is at the origin of the modification in swimming. The tensor $\bA_2$, oddly enough, conspires to give the correction of rescaling by the linear viscoelastic modulus.

\begin{figure}[t]
\centering
\includegraphics[width=0.5\textwidth]{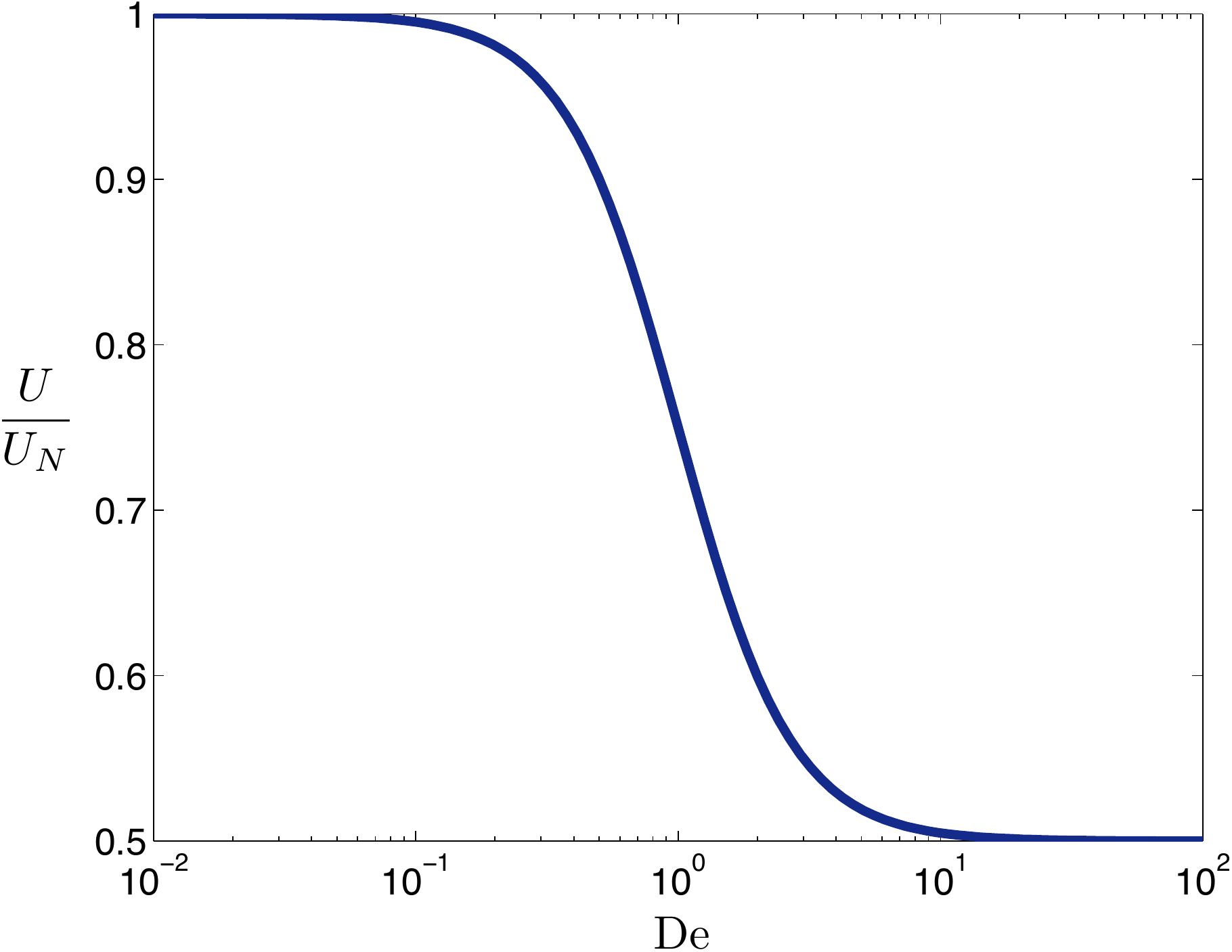}
\caption{The leading-order viscoelastic swimming speed scaled by the Newtonian solution, $U/U_N$, as a function of Deborah number, $\De=\omega\lambda_1$, for a Taylor swimming sheet in an Oldroyd-B fluid with $\beta=0.5$. The swimming speed is Newtonian, $U=U_N$, in the small Deborah number limit, $\De\rightarrow 0$, while $U=\beta U_N$ as $\De\rightarrow \infty$.}
\label{figspeed}
\end{figure}

For the  sinusoidal swimming sheet originally considered by Taylor,  the only nonzero coefficient is $c_1=-i/2$ leading to a swimming speed of $-\omega/2k$ in a Newtonian fluid \cite{taylor51}. The viscoelastic correction to this result, $\Rf[\eta^*(1)/\eta_0]$, was first derived in Ref.~\cite{lauga07}. The leading-order result is, written in a dimensional form,
\begin{align}\label{78}
U = -\frac{1}{2}a^2k\omega\frac{1+\beta\omega^2\lambda_1^2}{1+\omega^2\lambda_1^2}+\Of(\epsilon^4).
\end{align}
The swimming speed is no longer linear in the actuation frequency $\omega$, rather it is strictly decreasing with Deborah number from the Newtonian speed, $U=U_N$ as $\De\rightarrow 0$, to $U=\beta U_N$ as $\De\rightarrow\infty$, as illustrated in Fig.~\ref{figspeed}.

The rate at which a certain gait is performed now has a direct impact on the distance, $\Delta D$,  traveled over a period,
\begin{align}
\Delta D = -\pi a^2 k \frac{1+\beta\omega^2\lambda_1^2}{1+\omega^2\lambda_1^2},
\end{align}
clear evidence of the breakdown of Purcell's scallop theorem due to the presence of viscoelastic stresses \cite{fu09}.

Beyond kinematics, one might need to compute the power expended by the swimming motion. The energy dissipated in the fluid due to the actuation of the swimmer is given by an integration of $\btau_1:\bgammad_1$ over the whole fluid domain. Because the swimmer propagates periodic traveling waves, only the mean value is important  and it is given by 
\begin{align}
\lb<\btau_1:\bgammad_1\rb>=\sum_n \eta^*(n)\bgammad_1^{(n)}:\bgammad_1^{(-n)}.
\end{align}
We see again that the only difference between the viscoelastic dissipation here and that in a Newtonian fluid is the complex viscosity, $\eta^*(n)$, for each Fourier mode in comparison to $\eta_N$ in the Newtonian fluid. Note  that it matters, in a comparison of the dissipation, what we choose the Newtonian dissipation to be. As shown above as $\De \gg 1$ (or $n\gg 1$) $\Rf[\eta^*(n)/\eta_0]\approx \beta$ and a comparison to the energy dissipated in a Newtonian fluid with viscosity $\eta_N$ yields a factor of $\eta_0\beta/\eta_N$ which is less than one if $\eta_N = \eta_0$ (i.e~if we compare the dissipation in the complex fluid to that of a Newtonian fluid of the same viscosity). If the Newtonian fluid has the viscosity of the solvent in the Oldroyd-B fluid then $\eta_N=\beta\eta_0$ and we see by comparison that the viscoelastic energy dissipated is always higher, meaning the addition of elasticity to a solvent always increases energy dissipation. This effect is diminished for higher Fourier modes and at higher actuation rates and eventually, for $\De\gg 1$, only the solvent is dissipating energy as the network does not have time to flow \cite{lauga07}.

\subsection{Large-amplitude deformations}
One of the interesting debates in the  literature in the field concerns the question of whether some swimming gaits could see a swimming speed increase in viscoelastic fluids. It is obviously possible to construct a swimmer which has no net locomotion in a Newtonian fluid yet displays net motion in a non-Newtonian fluid and we discussed this strategy as a measurement technique to recover rheological properties of a fluid. Less obvious is the effect on swimming speed for a gait that already leads to nonzero net motion in a Newtonian fluid. We showed above that a swimming sheet of arbitrary geometry will always swim slower in a viscoelastic fluid. However, that result and the theory presented above rely on small gait amplitudes in order to take advantage of perturbation expansions. Clearly,  small-amplitude theory need not hold for gaits that are significantly straining the fluid.

Recent experiments by Liu \textit{et al.} \cite{liu11} showed a modest increase in swimming speed of a force-free (but not torque-free) rotating helix in a Boger fluid near $\De=1$ and showed that this enhancement is independent of end effects. In contrast, prior asymptotic analysis by Fu \textit{et al.} \cite{fu09} showed that, like the swimming sheet, the leading-order swimming speed (in a small-amplitude perturbation series) of a body propagating helical waves in an Oldroyd-B fluid is always slower than in a non-Newtonian fluid. This  discrepancy between the small-amplitude asymptotics and large-amplitude experiments  was resolved in numerical work by Spagnolie \textit{et al.} for a helix in an Oldroyd-B fluid \cite{spagnolie13}. They showed that there is a smooth transition between small-amplitude  hindered swimming and large-amplitude  enhanced locomotion near $\De=1$. The authors argued that a reason the speed enhancement occurs at an actuation rate that is on the same order as the relaxation rate of the fluid, $\omega^{-1} \approx \lambda_1$, may be because this is the time scale in which the flagellum revisits the viscoelastic wake it creates upon rotation. 
 
A question that one might ask is whether this mechanism translates to other unsteady swimming gaits. Numerical simulations by Teran \textit{et al.} \cite{teran10} showed that a finite two-dimensional swimmer, propagating deformation waves of increasing amplitude head to tail, sees an increase in swimming speed near $\De=1$ in an Oldroyd-B fluid with $\beta=1/2$. The authors rationalize that the increase in swimming speed results as a consequence of the highly strained fluid localized at the swimmers tail. Alternatively, in a set of experiments with the nematode \textit{C. Elegans} swimming in a Boger fluid, Shen and Arratia \cite{shen11} found that that non-Newtonian stresses strictly decrease the swimming speed. This nematode swims by propagating traveling waves with amplitudes which decay from head to tail. The functional dependence of the swimming speed on the Deborah number in their experiments \cite{shen11} is reminiscent of the systematic decay found  for a small-amplitude swimming sheet in an unbounded fluid \cite{lauga07}. Indeed numerical simulations documented later in this book (Ch.~10) show that if the gait of the nematode is reversed (reflection of the wavevector), yielding an increasing amplitude head to tail, then the nematode would experience a speed enhancement similar to the results presented by Teran \textit{et al}. Additionally, in studies of other swimming gaits, numerical simulations of potential squirmers \cite{zhu11} and pushers or pullers \cite{zhu12} all showed a speed decrease in a Giesekus fluid versus a Newtonian one.

In an effort to understand these large-amplitude results, we extend the small-amplitude result for Taylor's swimming sheet to higher order by deriving the next two orders of the perturbation series for the swimming velocity, $U=\epsilon^2U_2+\epsilon^4U_4+\epsilon^6U_6+\Of(\epsilon^8)$ (the perturbation series contains only even terms because of the $\epsilon\rightarrow -\epsilon$ symmetry). In a Newtonian fluid the first two terms were found by Taylor \cite{taylor51}, while the third was later derived by Drummond \cite{drummond66}. The series was recently resolved to arbitrarily high order by Sauzade \textit{et al.} \cite{sauzade11} who showed that the series converges only for small $\epsilon$, and then only slowly, but methods to accelerate convergence prove very effective enabling accurate prediction up of the swimming speed for order-one amplitudes.

The leading-order steady swimming speed in an Oldroyd-B fluid, $U_2$, was computed in Eq.~\eqref{78}. The next two nonzero orders in the asymptotic series, $U_4$ and $U_6$, can be found with a straightforward, but laborious, application of the formalism presented in this chapter. At fourth order we obtain analytically
\normalsize
\begin{align}
U_4=\omega k^{-1}&\frac{\left(1+\De^2 \beta\right)}{128 \left(1+\De^2\right)^3 \left(1+\De^2 \beta^2\right)}\times \nonumber\\
&\Bigg[76+50 \De^2+47 \De^4+\De^2 \left(102+29 \De^2\right) \beta
+\De^2 \left(76+45 \De^2+42 \De^4\right) \beta^2+\De^4 \left(107+34 \De^2\right) \beta^3\Bigg],
\end{align}
\normalsize
while at sixth order we find a much lengthier, but still entirely analytical formula (not shown).

\begin{figure}[t]
\centering
\includegraphics[width=0.5\textwidth]{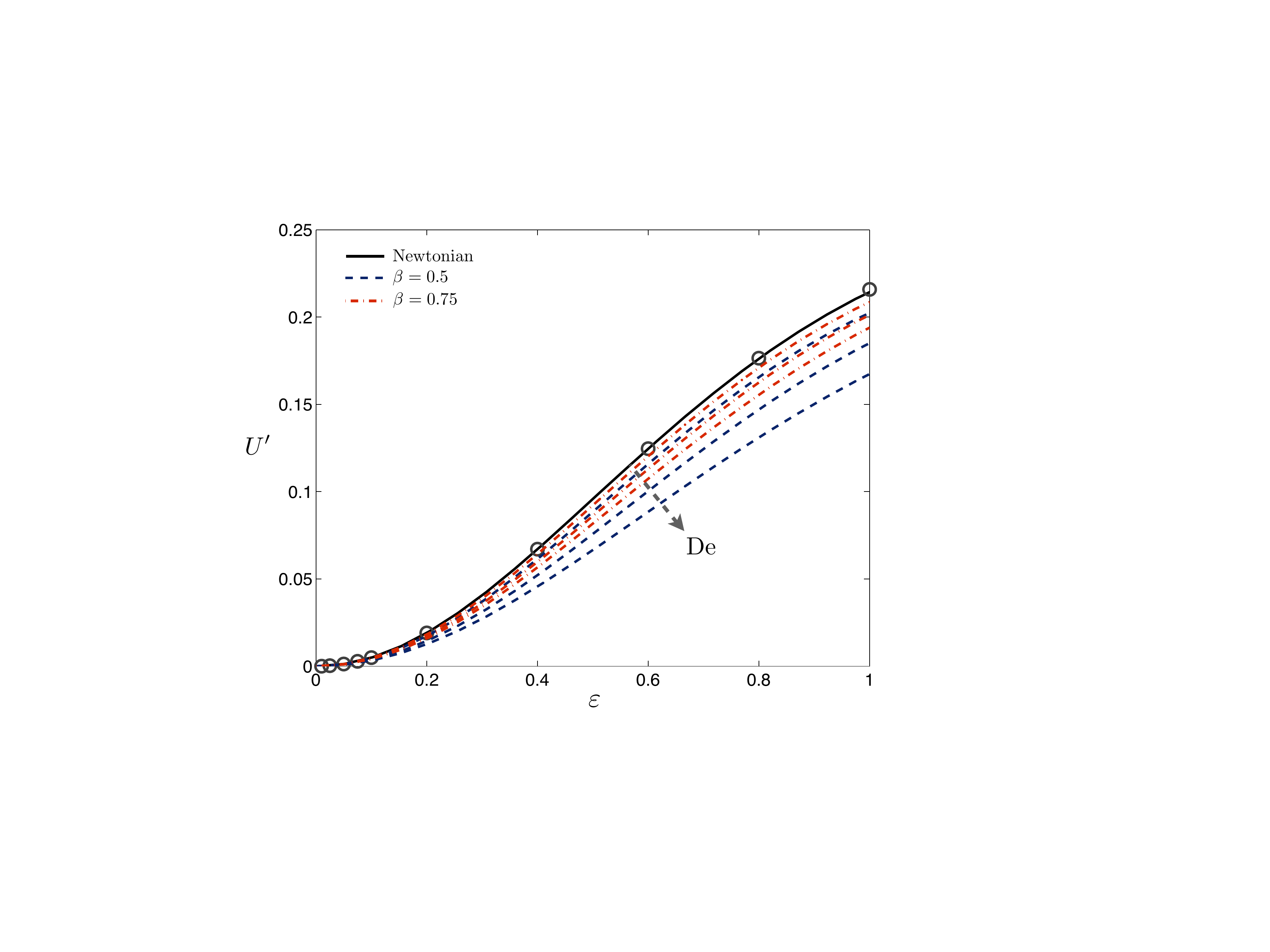}
\caption{Dimensionless swimming speed, $U'=U/\omega k^{-1}$, as a function of dimensionless amplitude, $\epsilon=ak$, for a Taylor swimming sheet in an Oldroyd-B fluid. Plotted is the $P^1_1(\epsilon)$ Pad\'e approximant of a three term perturbation series valid up to $\Of(\epsilon^6)$. The solid line shows the  Newtonian swimming speed, while the dashed line indicates the non-Newtonian result with $\beta = 0.5$ and dashed-dot is $\beta=0.75$, with Deborah numbers $\De = 0.5, 1, 1.5$. The circles are numerical results from boundary integral computation for $\De=0$. Viscoelasticity strictly decreases swimming speeds in this range of amplitudes.}
\label{pade}
\end{figure}

The swimming speed found using a conventional sum of the first three terms of the perturbation series for a Newtonian fluid is inaccurate past $\epsilon\approx 0.5$ but by using the method of Pad\'e approximants \cite{bender78}, with just three terms in the series, the $P^1_1(\epsilon)$ approximant is accurate up to $\epsilon \approx 1$ (within $1\%$ of the computational result determined using the  boundary integral method) \cite{sauzade11}. We follow the same tactic for the coefficients of the viscoelastic swimming speed based on the assumption that the Pad\'e $P^1_1(\epsilon)$ approximant is accurate for a larger range of amplitudes. In Fig. \ref{pade} we plot the swimming speed scaled by the wavespeed, $U'=U/\omega k^{-1}$, as a function of dimensionless amplitude $\epsilon = ak$. Under our modeling approach, we see that viscoelasticity serves to strictly hinder the swimming speed of the sheet even for dimensionless amplitudes on the order of $\epsilon = 1$. This result is in agreement with the experiments for the undulatory nematode but in contrast to numerical and experimental work on the propulsion of a rotating helix. 

Recent computational work  confirms that the Taylor sheet is indeed hindered at all amplitudes by the presence of viscoelastic stresses \cite{morozov13}. This was interpreted as due to  stagnation points in the stress near the peak and trough of the sheet, which act to retard the motion of the sheet. Exploiting this insight, sheets asymmetric about the horizontal axis, reminiscent of hyperactivated sperm flagella, may be constructed which see a speed increase in an Oldroyd-B fluid \cite{morozov13}. 

Recent experiments using a cylindrical variant of a Taylor sheet do show a speed increase in a Boger fluid (but not for a shear-thinning one) \cite{dasgupta13}. In these experiments a Couette-like device has a flexible inner cylinder which passes angular traveling waves of radial deformation while the outer cylinder rotates freely supported by low-friction bearings and hence at steady state yields an approximation of a swimming sheet near a wall.  A noticeable speed increase is observed if a Boger fluid is used instead of a Newtonian fluid for all wavespeeds but with an increasing difference for larger wavespeeds -- in stark contrast to the results for a planar sheet.

\subsection{Shear-dependent viscosity}

Many biological fluids through which microorganisms might swim, such as mucus, are not only viscoelastic but also have shear-dependent viscosity. Studies on the effects of a variable viscosity on the swimming speed of microorganisms, like those of elasticity, show  mixed results. Shen and Arratia's experiments with \textit{C. Elegans} show shear-thinning fluids have no noticeable effect on swimming speeds \cite{shen11}. In contrast, experiments on the cylindrical Taylor sheet show a marked decrease in the swimming speed \cite{dasgupta13}. 

A recent theoretical study on the Taylor swimming sheet in a Carreau  fluid was undertaken by V\'elez-Cordero and Lauga \cite{velezcordero13}. A Carreau fluid\index{Carreau fluid} has a shear-rate-dependent deviatoric stress given by
\begin{align}\label{Carreau}
\btau = \eta_0\lb[1+\lambda_t^2\lb|\gammad\rb|^2\rb]^{N} \bgammad,
\end{align}
with $\lb|\gammad\rb|^2 = \bgammad:\bgammad/2$ and $N=(n-1)/2$ where $n$ is the so-called power-law index for the fluid and $\lambda_t$ is the relaxation time of fluid. Physically, we see from Eq.~\eqref{Carreau} that at high shear rates, the typical shear stress scales as $\tau \propto \dot\gamma^n$ and thus the number $n$ is the  power-law dependence in the stress/shear rate relationship  at high shear rates.   If the typical shear rate in the flow scales as $\omega$ then the Carreau number $\Cu = \omega\lambda_t$ is a dimensionless measure of the extent to which the fluid viscosity is altered; if $\Cu \ll 1$ the fluid behaves as Newtonian with (zero-shear-rate) viscosity $\eta_0$. 

In the small-amplitude study of Ref.~\cite{velezcordero13} it was found that a shear-thinning fluid has actually no effect on the swimming speed of the sheet if it deforms inextensibly. If instead the motion of the material is extensible, then there is an additional higher order non-Newtonian contribution, $U - U_N \sim \pm \epsilon^4 N\Cu^2$, with a sign which depends on the details of the waving kinematics. Recent numerical work at high amplitude and for finite swimmers by Montenegro-Johnson \textit{et al.} confirms that the (often weak) effects of a shear-thinning fluid depend on the gait of the swimmer with examples of both faster and slower swimming given for a variety of model swimmers \cite{montenegro12, montenegro13}.

In addition, we note that the swimming sheet model has also been used to address swimming near a wall in shear-rate-dependent and yield-stress fluids  \cite{chan05, lauga06, balmforth10}. In these studies, it was assumed that the separation between the sheet and the wall was much smaller than the wavelength of the sheet, $hk\ll 1$, thus taking advantage of the long-wavelength  lubrication approximation.

\subsection{Prescribed forcing}
Instead of imposing the wave kinematics, an alternative modeling approach to the problem of locomotion consists in  prescribing the internal forcing which, through a dynamic balance,  leads to the deformation of the body. Fu and Powers investigated theoretically the effects of viscoelasticity on the shape of a beating eukaryotic flagellum  \cite{fu07}. Flagellar beat patterns are determined by an interplay between the mechanical properties of the flagellum, the internal action from the  dynein motor proteins which produce active bending moment,  and the   hydrodynamic forces \cite{riedelkruse07}. Using a simplified sliding filament model for a sperm flagellum, the investigation in Ref.~\cite{fu07}  showed that the introduction of viscoelasticity into the fluid can dramatically affect the  shape of the waving flagellum.

One approach to model the impact of this change in kinematics  on locomotion is  to consider the body of a swimmer to be composed of a repeated series of simple shapes,  such as spheres. Najafi and Golestanian proposed the simplest such swimmer which can locomote in Newtonian fluid \cite{najafi04}, consisting of  three equal co-linear spheres with prescribed, periodically varying, relative displacements (as if connected by hydrodynamically invisible rods of time-varying lengths). A system of two such spheres cannot self-propel  in a Newtonian fluid as it has but a single degree of freedom, but with three spheres one can produce time-irreversible motion and swimming \cite{najafi04}.  

Curtis and Gaffney recently endeavored to determine the motion of the three-sphere swimmer in an Oldroyd-B fluid \cite{curtis13}. In general the superposition of far-field singularity solutions, which is the standard method to solve the N-sphere Newtonian problem,  is not possible in a viscoelastic fluid due to the nonlinearities in the  governing equations. However, if the flow field is resolved as a perturbation expansion in small-amplitude disturbances then the constitutive relation can be linearized order-by-order. After expanding to quadratic order in small-amplitude motion (relative to the sphere radii),  and for prescribed kinematics, the net displacement over a period is found to be identical to that of a Newtonian fluid \cite{curtis13}. This surprising result, at odds with the swimming sheet result, might be a consequence of the far-field approximation. However, the simplicity of this model allows for a straightforward implementation of prescribed-forcing case, rather than prescribed kinematics. In this case, the relative forces exerted by each sphere on its neighbor are prescribed and the gait and swimming kinematics are then solved for. For prescribed forcing the time-averaged swimming speed $U$ differs from the Newtonian swimming speed $U_N$ by a factor 
\begin{align}
\frac{U}{U_N} =\frac{1+\De^2}{1+\beta^2\De^2},
\end{align}
which is always greater than one since $\beta \le 1$. Prescribing the internal forcing of a swimmer rather than its kinematics  leads to a qualitatively different response and a speed increase ensues. It will be interesting to extend this approach in the future to more realistic models of swimming cells. 

We close by also noting that numerical simulations performed using the immersed boundary method by Chrispell \textit{et al.} \cite{chrispell13} involve specifying preferred waving kinematics with the true kinematics then resolved as a balance between elastic and fluid forces (in a sense, a hybrid between specifying kinematics and specifying the internal forces).

\subsection{Two-fluid models}\index{Two-fluid models}

\begin{figure}[t]
\centering
\includegraphics[width=.75\textwidth]{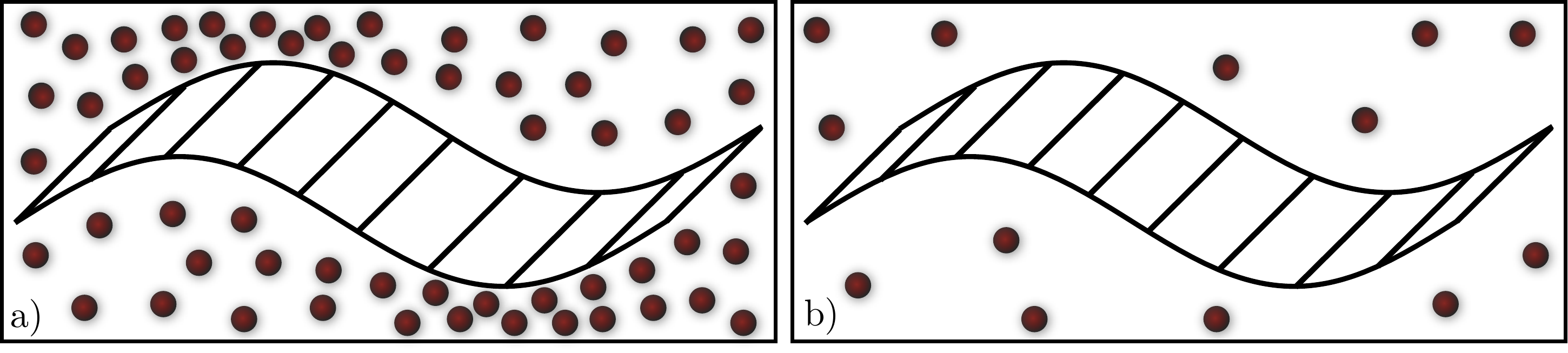}
\caption{Schematic  representation of a Taylor swimming sheet in a two-fluid model fluid \cite{fu10}. (a) in a dense network the sheet interacts directly with the elastic network; (b) if the network is dilute the interactions between the sheet and the network are  mediated through the fluid.}
\label{two-fluid}
\end{figure}

Microorganisms may swim in environments where heterogeneities in the fluid exist on length scales similar to the  swimmers themselves.  Berg and Turner found propulsion enhancement in gel-like environments (methylcellulose solutions) where the solute forms a loose, quasi-rigid network \cite{berg79}. They postulated the microorganisms could then push directly on this network and that a helical flagellum would move as would a corkscrew through a cork, with less  circumferential slip than in a Newtonian fluid. In order to properly capture this behavior the material should possess  a nonzero zero-frequency elastic shear modulus, and   the swimmer must be able to slip past the solid medium.

Fu \textit{et al.} used a two-fluid model to explore the effects of a cross-linked network on the swimming speed of Taylor's waving sheet \cite{fu10} (illustrated in Fig. \ref{two-fluid}). In this model a drag force density, proportional to the relative local velocity,
\begin{equation}
\bf_d = \Gamma\left(\frac{\d}{\d t}\bX-\bu\right), 
\end{equation}
couples the elastic network (displacement field $\bX$) to the Newtonian solvent (velocity field $\bu$), 
\begin{equation}
\bnabla\cdot\bsigma^n = \bf_d, \quad\bnabla\cdot\bsigma^s = -\bf_d,
\end{equation}
where $n$ indicates the network and $s$ indicates the solvent. The friction coefficient $\Gamma$ introduces an intrinsic length scale into the problem typically known as the screening length, $l_s = \sqrt{\eta/\Gamma}$. The ratio of the screening length to the physical length scale of the sheet, $kl_s$, is a dimensionless measure of the interaction between the fluid flow and the network and as expected if  $kl_s \gg 1$ then the sheet effectively sees only a Newtonian fluid.

If the network is dense as illustrated in Fig. \ref{two-fluid}a, the sheet comes into direct contact with the network which is then  deformed  while permitting tangential slip via a Navier friction law
\begin{equation}
\bt\cdot\bsigma^n(y=y_1)\cdot\bn=\chi\bt\cdot\lb.\left(\frac{\d}{\d t}\bX-\bu\right)\rb|_{y=y_1},
\end{equation}
 with slip coefficient $\chi$. If the network is dilute as illustrated in Fig. \ref{two-fluid}b, then no contact is made and hence no traction is applied to the sheet, $\bsigma^n(y=y_1)\cdot\bn=\bzero$.   

The rigidity of the network has a dramatic effect on the swimming speed,  and stiff networks enhance swimming speed compared to a Newtonian fluid while compliant networks retard swimming speed, regardless of whether the network contact is direct or solvent mediated \cite{fu10}. If the network is considered immobile then the model above reduces to a Brinkman fluid \cite{brinkman49}. In this limit the swimming speed is seen to be systematically enhanced by a factor
\begin{align}
\frac{U}{U_N}=\sqrt{1+\Gamma/(\eta k^2)},
\end{align}
as shown by Leshansky \cite{leshansky09}. As expected if we take $kl_s \gg 1$ then the swimming speed reduces to that in a Newtonian fluid.

A two-fluid model was also used by Du \textit{et al.} to study two intermixed Newtonian fluids of different viscosities \cite{du12}. In this case it was shown that swimming in a mixture of two Newtonian fluids is always slower and less efficient than swimming in a single viscous fluid.

\subsection{Collective effects}

When microorganisms are swimming  close to one another, they may interact hydrodynamically. Namely the presence of a  flow field created by one organism affects the dynamics of nearby swimmers, and vice versa. Because the decay of a low-Reynolds number flow field is often long tailed, hydrodynamic interactions may significantly alter collective dynamics. For instance, dense suspensions of microswimmers display transient ordered flow structures  with much larger length and velocity scales than the organisms themselves \cite{dombrowski04}. Hydrodynamic interactions also enable synchronous flagellar beating for the  alga \textit{Chlamydomonas}, synchrony that then facilitates directed locomotion \cite{polin09}.

Non-Newtonian fluids affect not only how microorganisms self-propel individually, but also impact  hydrodynamic interactions between microogranisms in close proximity. A striking example of hydrodynamic synchronization occurs in the phase locking of two (or several) spermatozoa flagella when they are near one another \cite{woolley09}. This synchronization leads to a speed increase of the group of cells and thereby a competitive advantage. 

Taylor first attempted to model synchronization in the Newtonian case by solving for the hydrodynamic interaction between two sinusoidal sheets. He found that energy dissipation in the fluid is minimized if the two sheets are oscillating with no phase difference \cite{taylor51}. However,  in this  symmetric  sinusoidal setup, there can be no evolution of the phase from an arbitrary initial condition, due to the kinematic reversibility of the Stokes flow field equations \cite{elfring09, elfring11}. If two such sheets are swimming in  a viscoelastic fluid, such as those present along the path through the female reproductive system, then kinematic reversibility no longer constrains the dynamics. Two Taylor sheets are found to systematically  synchronize to an in-phase conformation in an Oldroyd-B fluid  \cite{elfring10}. 

Here again, the reciprocal theorem may be used to study the interactions between two general sheets in a complex fluid. Taking into account the force-free motion of both sheets we find that in an Oldroyd-B fluid the phase, $\phi$, between two sheets evolves in time as
\begin{align}
\frac{\d\phi}{\d t} &=-\epsilon^2\omega\sum_n 2 c_nc_n^\dagger n  \sin (n \phi ) C(nh)\frac{G^*(n)}{\eta_0\omega},
\end{align}
where $C(x) = (x\sinh x+x^2 \cosh x)/(\sinh ^2x-x^2)$. The rate of sheet synchronization is thus dependent on the elastic modulus of the of the fluid, $\Rf[G^*/\eta_0\omega]=n^2(1-\beta)\De/(1+n^2\De^2)$ where $G^*(n) = in\omega\eta^*(n)$, rather than the viscous modulus which affects the collective locomotion speed. Clearly, in  a Newtonian fluid ($\De=0$) there is no evolution of phase in time while the addition of viscoelastic forces ($\De\ne 0$ and $\beta \ne 1$) leads to the evolution of all initial configurations to an in-phase state, $\phi=0$. This is similar to the synchronization of elastic sheets in a Newtonian fluid, whereby deformations due to fluid-structure interactions lead to an evolution of phase \cite{elfring11b}. Similar results were recently demonstrated numerically \cite{chrispell13}, a study that also addressed the transient evolution of stress from an initial condition.

A variety of numerical and analytical studies have examined the evolution of suspensions of swimmers in Newtonian fluids, as reviewed for example by  Saintillan and Shelley in the following chapter of this book (Ch.~9). Recent studies have adapted the mean-field theories for suspensions in Newtonian fluids \cite{saintillan08,saintillan08b,hohenegger10,saintillan13} to quantify the impact of the addition of viscoelasticity \cite{underhill11,underhill13}. Specifically a polymeric stress is incorporated into the mean-field description of the flow which is forced by a configurational average of the force dipoles exerted by the swimmers on the fluid. Under these conditions the stability of an isotropic suspension is qualitatively unchanged for up $\De \approx \Of(1)$ while showing significant variation for larger Deborah number.

We finally note that the addition of swimmers (more generally, a suspension of active particles) modifies the apparent rheology of the fluid when viewed from a continuum perspective \cite{hatwalne04}. That swimmers should modify the bulk rheology of a system is already clear by simply looking at the passive analogue. As discussed earlier in this book (Ch.~3), a dilute suspension of rigid spheres leads to the famous Einstein correction to viscosity \cite{einstein56}, while the addition of Brownian motion and weakly anisotropic particles leads to non-Newtonian rheological features \cite{leal72}. Passive rod-shaped Brownian particles tend to align with imposed shear on average and hence a suspension of rods, while possessing a larger zero-shear rate, displays shear thinning at higher shear rates. The preferred alignment of the swimmers with the shear direction then leads, in the active case, to additional stresslets imposed on the fluid and further impact the rheological properties.

\section{Perspective}\label{sec:5}
Despite the fact that the first studies on the locomotion of microorganisms in non-Newtonian fluids are over  thirty years old, the bulk of the field has developed very recently. In spite of recent progress in developing a theory to describe the effects which non-Newtonian fluids have on swimming kinematics, many subtleties remain to be parsed out. What we do know, is that there is no simple answer as to the impact of viscoelasticity on locomotion. Changes in the gait of a swimmer seem to lead to drastic changes in the non-Newtonian effects. For example, when propagating helical waves, a swimmer is retarded by a viscoelastic (Boger) fluid if the helical amplitude is small but enhanced if the helical amplitude is large. In contrast, for swimmers propagating sinusoidal undulatory waves, elastic stresses only serve to slow it down regardless of the amplitude of the motion, yet if the amplitude is increasing tip to tail on finite swimmers, a speed increase may yet be obtained. 

In general complex fluids are both shear-dependent and viscoelastic, and biological swimming gaits show a great deal of diversity. Developing an understanding of the differences which arise amongst various gaits and the non-Newtonian response of a particular fluid will lead to insight into the swimming strategies observed in different natural environments. In particular, future work should help shed light on  how organisms  passively or actively modulate their behavior to cope with complex stresses. This understanding may also lead to more effective designs of artificial microswimmers  for use in biological environments such as in therapeutic delivery devices. As is often the case in physics, progress in this field so far has been achieved  through careful analysis of the locomotion of  simple model swimmers which then raises questions to be addressed by numerical simulation and experiment and we hope that in the future, work will be also fueled by novel biological experimental insights.

\section*{Acknowledgements}
GE gratefully acknowledges funding from the Natural Science and Engineering Research Council of Canada while EL thanks the European Union (through a CIG Grant) for partial support. 

\bibliographystyle{unsrt}

\bibliography{swimming}

\end{document}

%% file: definitions.tex
\usepackage{bm}
\def\boldsymbol{\bm}

\def \t{\tensorsym}
\def \lb{\left}
\def \rb{\right}

\def \d{\,\text{d}}

\newcommand{\upc}[1]{\overset{\triangledown}{#1}}

\def \etah{\hat{\eta}}
\def \bgamma{\boldsymbol{\gamma}}
\def \bgammadt{\tilde{\dot{\boldsymbol{\gamma}}}}
\def \bgammad{\dot{\boldsymbol{\gamma}}}
\def \bgammadh{\hat{{\dot{\bgamma}}}}
\def \gammad{{\dot{\gamma}}}
\def \bnabla{\boldsymbol{\nabla}}
\def \bsigma{\boldsymbol{\sigma}}
\def \bsigmah{\hat{\boldsymbol{\sigma}}}
\def \btau{\boldsymbol{\tau}}
\def \btaut{\tilde{\boldsymbol{\tau}}}
\def \btauh{\hat{\boldsymbol{\tau}}}
\def \bOmega{\boldsymbol{\Omega}}
\def \bOmegah{\hat{\boldsymbol{\Omega}}}
\def \bTheta{\boldsymbol{\Theta}}
\def \bXi{\boldsymbol{\Xi}}

\def \Ar{\text{Ar}}
\def \Cu{\text{Cu}}
\def \De{\text{De}}
\def \Re{\text{Re}}

\def \bA{\mathbf{A}}
\def \bAt{\tilde{\mathbf{A}}}
\def \bB{\mathbf{B}}
\def \bBt{\tilde{\mathbf{B}}}
\def \bC{\mathbf{C}}
\def \bF{\mathbf{F}}
\def \bG{\mathbf{G}}
\def \bI{\mathbf{I}}
\def \bL{\mathbf{L}}
\def \bN{\mathbf{N}}
\def \bR{\mathbf{R}}

\def \bU{\mathbf{U}}
\def \bX{\mathbf{X}}

\def \Rh{\hat{R}}
\def \Uh{\hat{U}}

\def \bFh{\hat{\mathbf{F}}}
\def \bLh{\hat{\mathbf{L}}}
\def \bRh{\hat{\mathbf{R}}}
\def \bUh{\hat{\mathbf{U}}}

\def \be{\mathbf{e}}
\def \bf{\mathbf{f}}
\def \bn{\mathbf{n}}
\def \br{\mathbf{r}}
\def \bt{\mathbf{t}}
\def \bu{\mathbf{u}}
\def \bx{\mathbf{x}}

\def \fd{\dot{f}}

\def \bzero{\mathbf{0}}

\def \buh{\hat{\bu}}

\def \Af{\mathcal{A}}
\def \Bf{\mathcal{B}}
\def \Lf{\mathcal{L}}

\def \Of{\mathcal{O}}
\def \Rf{\mathcal{R}}

\def \tzero{\mathsf{\t 0}}

\def \tF{\mathsf{\t F}}

\def \tR{\mathsf{\t R}}
\def \tU{\mathsf{\t U}}

\def \tFh{\mathsf{\t{\hat{F}}}}
\def \tGh{\mathsf{\t{\hat{G}}}}
\def \tRh{\mathsf{\t{\hat{R}}}}
\def \tTh{\mathsf{\t{\hat{T}}}}
\def \tUh{\mathsf{\t{\hat{U}}}}